\documentclass[twocolumn,showpacs,preprintnumbers,amsmath,amssymb,prb,superscriptaddress]{revtex4-1}
\usepackage{mathrsfs}
\usepackage{graphicx}
\usepackage{dcolumn}
\usepackage{bm}
\usepackage{amsmath}
\usepackage{amsfonts}
\usepackage{color}

\begin{document}

\title{Quantum phase transition of chiral Majorana fermion in the presence of disorder}
\author{Biao Lian}
\affiliation{Department of Physics, McCullough Building, Stanford University, Stanford, California 94305-4045, USA}
\affiliation{Princeton Center for Theoretical Science, Princeton University, Princeton, New Jersey 08544, USA}
\author{Jing Wang}
\affiliation{State Key Laboratory of Surface Physics, Department of Physics, Fudan University, Shanghai 200433, China}
\affiliation{Collaborative Innovation Center of Advanced Microstructures, Nanjing 210093, China}
\author{Xiao-Qi Sun}
\affiliation{Department of Physics, McCullough Building, Stanford University, Stanford, California 94305-4045, USA}
\author{Abolhassan Vaezi}
\affiliation{Department of Physics, McCullough Building, Stanford University, Stanford, California 94305-4045, USA}
\author{Shou-Cheng Zhang}
\affiliation{Department of Physics, McCullough Building, Stanford University, Stanford, California 94305-4045, USA}

\begin{abstract}
We study the quantum phase transitions of a disordered two-dimensional quantum anomalous Hall insulator with $s$-wave superconducting proximity, which are governed by the percolation theory of chiral Majorana fermions. Based on symmetry arguments and a renormalization group analysis, we show there are generically two phase transitions from Bogoliubov-de Gennes Chern number $\mathcal{N}=0$ to $\mathcal{N}=1$ ($p+ip$ chiral topological superconductor) and then to $\mathcal{N}=2$, in agreement with the conclusion from the band theory without disorders. Further, we discuss the critical scaling behavior of the $e^2/2h$ conductance half plateau induced by $\mathcal{N}=1$ chiral topological superconductor recently observed in the experiment. In particular, we
compare the critical behavior of the half plateau induced by topological superconductor with that predicted recently by alternative explanations of the half plateau, and show that they can be distinguished in experiments.
\end{abstract}

\date{\today}

\maketitle

\section{Introduction}
The search for two-dimensional (2D) $p+ip$ chiral topological superconductor (TSC) with a Bogoliubov de-Gennes (BdG) Chern number $\mathcal{N}=1$ has attracted extensive theoretical and experimental interest in the past decades \cite{fu2008,fu2009a,sato2009a,sau2010,alicea2010,raghu2010,tanaka2009a,akhmerov2009,
mackenzie2003,lutchyn2010,qi2011,wangqh2013,law2009,he2014,xu2015,qi2010b,chung2011,wang2015a,he2017}. As one of the simplest topological phases of matter, it gives rise to chiral Majorana fermions on the edge which are anti-particles of themselves, and Majorana zero modes in $\pi$ flux vortices in the bulk which obey non-Abelian statistics and have a potential application in topological quantum computations \cite{moore1991,wen1991,read2000,ivanov2001,nayak2008,kitaev2003}. In particular, it has been proposed that the $\mathcal{N}=1$ chiral TSC can be realized in a heterostructure of a 2D quantum anomalous Hall insulator (QAHI) ferromagnetic thin film and an s-wave superconductor (SC) \cite{qi2010b,chung2011,wang2015a}. In the vicinity of the phase transition of the thin film from QAHI to normal insulator (NI), such a structure is predicted to exhibit a half quantized two-terminal conductance plateau of value $e^2/2h$ in between two integer plateaus $0$ and $e^2/h$, which signals the occurrence of chiral Majorana edge fermion and the realization of $\mathcal{N}=1$ chiral TSC \cite{chung2011,wang2015a}. In accordance with the theoretical prediction, such a half plateau is observed in the experiment of He \emph{et al.} \cite{he2017}, where heterostructures of Cr doped thin film (Bi,Sb)$_2$Te$_3$ QAHI \cite{chang2013b,checkelsky2014,kou2014,chang2015} and Nb superconductor are adopted, which is the first quantized feature for chiral Majorana fermions in condensed matter experiments.

The original theory of the half plateau from chiral Majorana fermion is formulated in a homogeneous system \cite{qi2010b,chung2011,wang2015a}. However, the half plateau in the experiment occurs in the magnetization flipping stage of the thin film \cite{he2017}, which is expected to be extremely inhomogeneous. Alternative explanations of the half plateau under strong disorders have been raised \cite{ji2017,huang2017}, questioning the experimental realization of $p+ip$ chiral TSC. In this paper, we employ the percolation theory of chiral Majorana fermion \cite{kramer2005,chalker2001} to study the effects of disorder on the chiral TSC phase. Based on both symmetry arguments and renormalization group (RG) analysis, we show that the $\mathcal{N}=1$ chiral TSC phase in such a system is robust under disorders. Furthermore, we study the critical behaviors of the transitions from the chiral TSC induced half plateau $e^2/2h$ to $0$ and that from $e^2/2h$ to $e^2/h$, and show they are quite distinct from those in the alternative explanations \cite{ji2017,huang2017}. In particular, the $e^2/2h$ to $0$ transition belongs to the A symmetry class with a localization length critical exponent $\nu_A\approx7/3$, while the $e^2/2h$ to $e^2/h$ transition belongs to the D symmetry class with a localization length critical exponent $\nu_D\approx1$, which lead to different size and temperature scalings that can be tested in current and future experiments.

The paper is organized as follows. In Sec. \ref{Hamiltonian} we briefly review the rise of $p+ip$ chiral TSC in a homogeneous QAHI system proximity coupled with a uniform $s$-wave SC, and the mechanism of half conductance plateau due to the single chiral Majorana edge state of the TSC. In Sec. \ref{Network} we illustrate the chiral Majorana fermion percolation picture of the system in the presence of disorders, and formulate it as a particle-hole symmetric network model in the D symmetry class without time-reversal and spin-rotational symmetries~\cite{kramer2005}. Via an RG analysis, we show the $\mathcal{N}=1$ chiral TSC phase is stable against disorders. In Sec. \ref{Critical} we discuss the critical size and temperature dependent scaling of the half plateau induced by chiral TSC and its transition to neighbouring integer plateaus, and show they are completely different from the critical behaviors predicted in alternative explanations \cite{ji2017,huang2017}, which are testable in current and future experiments. At last, we draw our conclusions in Sec. \ref{Conclusion}.

\section{TSC from a homogeneous QAHI}\label{Hamiltonian}

In the absence of disorders, the low energy BdG Hamiltonian of a 2D QAHI in the vicinity the Hall plateau transition from $\sigma_{xy}=0$ to $\sigma_{xy}=e^2/h$  under the proximity of an $s$-wave SC is as follows:
\begin{equation}
H_{\text{BdG}}(\mathbf{k})=\left(\begin{array}{cc}h(\mathbf{k})-\mu&i\Delta\sigma_y\\ -i\Delta^*\sigma_y&-h^*(-\mathbf{k})+\mu\end{array}\right)\ ,
\end{equation}
where $\sigma_{x,y,z}$ are Pauli matrices, and
\[
h(\mathbf{k})=(m-D\mathbf{k}^2)\sigma_z+ Ak_x\sigma_x+Ak_y\sigma_y\ ,
\]
with $D>0$. The Nambu basis of the BdG Hamiltonian is $(c_{\uparrow,\mathbf{k}},c_{\downarrow,\mathbf{k}},c_{\uparrow,-\mathbf{k}}^\dag,c_{\downarrow,-\mathbf{k}}^\dag)$, where $c_{s,\mathbf{k}}$ and $c_{s,\mathbf{k}}^\dag$ are the electron annihilation and creation operators with spin $s$ and momentum $\mathbf{k}$. In the magnetic TI thin film, $A$ is the spin-orbital coupling strength, $\mu$ is the chemical potential, $\Delta$ is the s-wave proximity pairing amplitude, while the Dirac mass $m=\lambda-\lambda_0$ is half of the QAHI gap, where $\lambda$ is the ferromagnetic exchange field and $\lambda_0$ is a constant. The BdG Chern number $\mathcal{N}$ of such a superconductor is given by
\begin{equation}\label{BdGN}
\mathcal{N}=\begin{cases}
0\ ,\quad (m<-\sqrt{\mu^2+|\Delta|^2})\\
1\ ,\quad (|m|<\sqrt{\mu^2+|\Delta|^2})\\
2\ ,\quad\  (m>\sqrt{\mu^2+|\Delta|^2})
\end{cases}
\end{equation}
The $\mathcal{N}=0$ phase and $\mathcal{N}=2$ phase are topologically equivalent to the NI and QAHI phases, respectively, while the $\mathcal{N}=1$ phase is the $p+ip$ chiral TSC phase with a single chiral Majorana state on its edge. All the three phases have a finite gap for quasiparticles. Therefore, the system undergoes two phase transitions from $\mathcal{N}=2$ to $\mathcal{N}=0$ as a function of the exchange field $\lambda$ when $\Delta\neq0$ \cite{qi2010b}.

When an $s$-wave SC is in proximity with a QAHI in the way as shown in Fig. \ref{domain}(a), the system becomes a QAHI/TSC/QAHI junction for $0<m<\sqrt{\mu^2+|\Delta|^2}$, where the TSC has $\mathcal{N}=1$, and the chemical potential $\mu$ is in the QAHI gap. In the homogeneous case, the two-terminal conductance of the junction can be shown \cite{chung2011,wang2015a} to be quantized into one half of the quantum conductance $\sigma_{12}=e^2/2h$.
This can be understood as follows. As shown in Fig. \ref{domain}(a), each of the charged chiral edge states $\psi_i$ of the left and right QAHI regions split into two chiral Majorana edge states $\gamma_i$ at the boundaries of the TSC ($1\le i\le4$). At zero energy, all the chiral Majorana fermions $\gamma_i$ are charge neutral, and they are related to the charged chiral fermions $\psi_i$ via the relation $\psi_i=\gamma_i+i\gamma_{i+1}$, where $\gamma_5$ is identified with $-\gamma_1$. When an electron is injected into chiral edge state $\psi_1$, it has equal probabilities to become either an electron or a hole in either chiral states $\psi_2$ or $\psi_4$. This gives $r=r_A=t=t_A=1/4$, where $r$, $r_A$, $t$, $t_A$ are the normal reflection, Andreev reflection, normal transmission and Andreev transmission probabilities of the junction, respectively. According to the generalized Landauer-B\"uttiker formula \cite{anantram1996,entin2008}, the two terminal conductance is given by $\sigma_{12}=(t+r_A)e^2/h=e^2/2h$. Therefore, the system will exhibit a $\sigma_{12}=e^2/2h$ half quantized plateau for $0<m<\sqrt{\mu^2+|\Delta|^2}$. In contrast, when $m>\sqrt{\mu^2+|\Delta|^2}$, the middle region is in the same phase as QAHI, so $\sigma_{12}\rightarrow e^2/h$. While when $m<0$, the left and right regions of the junction become NIs, and one would have $\sigma_{12}\rightarrow 0$.

\section{Stability of TSC against disorder}\label{Network}

The experiment of He \emph{et al.} \cite{he2017} observed the half plateau as predicted above in the Cr doped (Bi,Sb)$_2$Te$_3$ thin film QAHI with proximity from a Nb superconductor, which provides a strong evidence for the realization of chiral TSC. The Dirac mass $m=\lambda-\lambda_0$ of the ferromagnetic QAHI thin film is controlled by the magnetic field $B$. When the system is deep in the QAHI phase, the film is a ferromagnetic single domain with a homogeneous exchange field $\lambda>\lambda_0$. By adding a magnetic field $B$ opposite to the magnetization, the system is driven into a disordered multi-domain configuration, while the spatial average value of the exchange field $\langle \lambda (\mathbf{r})\rangle$ is gradually reduced. To a good approximation, we have $\langle\lambda\rangle-\lambda_0=\langle m\rangle \propto B-B_0$ for small $m$, where $B_0$ is the magnetic field at which the average Dirac mass $\langle m\rangle=0$. The multi-domain structure during the decrease of $\langle\lambda\rangle$ leads to enormous inhomogeneities into the system, so the phase transitions can no longer be well understood within the homogeneous band theory above.

\begin{figure}[tbp]
  \centering
  \includegraphics[width=3.4in]{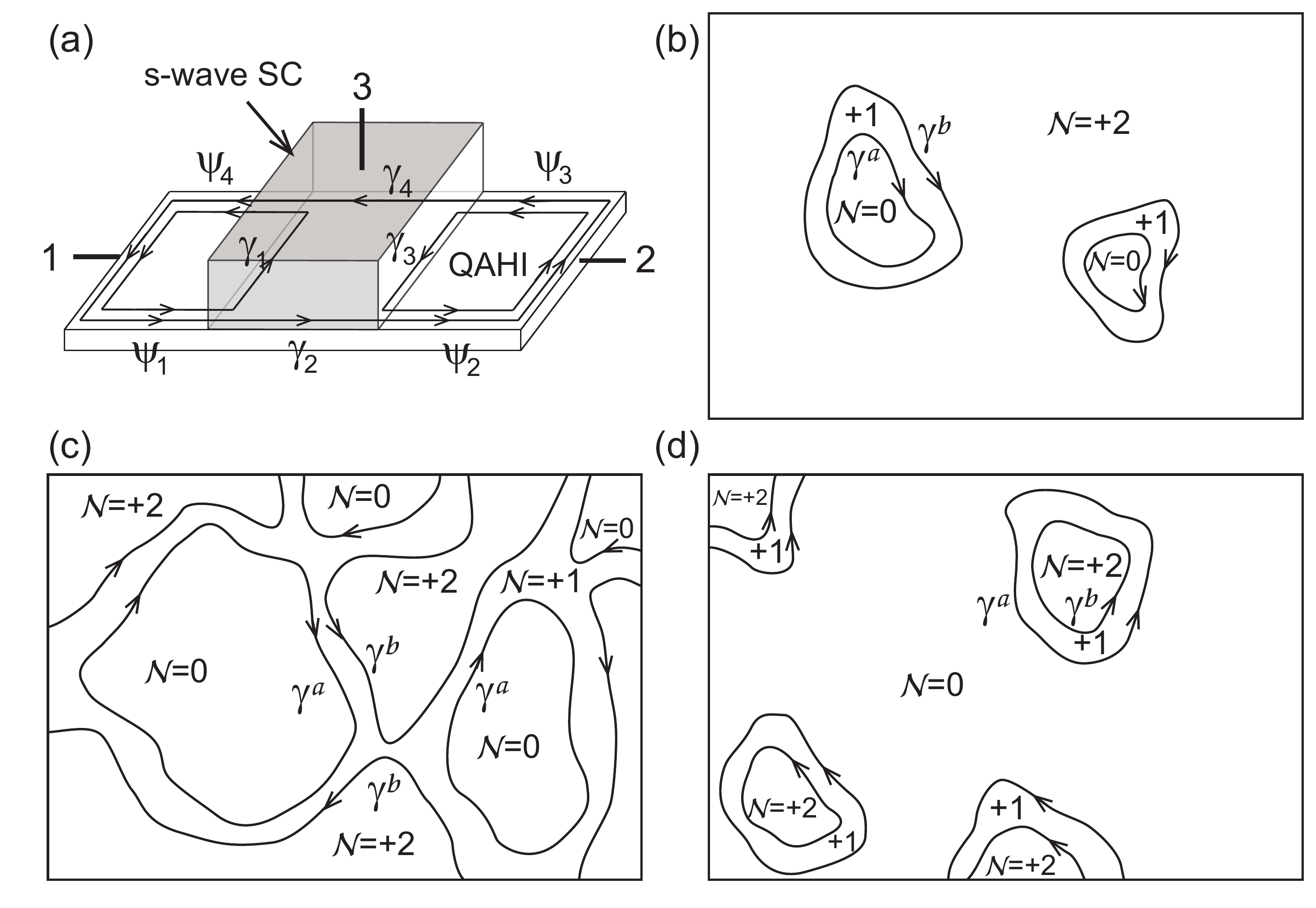}
  \caption{(a) Illustration of the QAHI/TSC/QAHI junction, where a QAHI thin film is in proximity with an s-wave SC on top in the middle region. (b)-(d) Evolution of BdG Chern number $\mathcal{N}=0,1,2$ regions during the domain flipping process, where each of the two chiral Majorana states $\gamma_a$ and $\gamma_b$ undergoes a delocalization transition. (b), (c) and (d) show configurations where $\mathcal{N}=2,1$ and $0$ regions are extended, respectively.}\label{domain}
\end{figure}

Instead, the random domain structure requires the knowledge of the percolation theory for understanding the phase transitions \cite{chalker1988,wang2014a}. In general, all the three parameters, the Dirac mass $m(\mathbf{r})$, the chemical potential $\mu(\mathbf{r})$ and the pairing potential $\Delta(\mathbf{r})$ are spatially nonuniform. Depending on the local values of $m,\mu$ and $\Delta$, the space can be divided into three kinds of regions according to Eq. (\ref{BdGN}) with BdG Chern numbers $\mathcal{N}=0,1,2$, respectively. As shown in Fig. \ref{domain}(b)-(d), there will be a charge neutral chiral Majorana edge state on each edge in between two regions with $\mathcal{N}$ differing by $1$. We shall denote the chiral Majorana fermion between $\mathcal{N}=0$ and $\mathcal{N}=1$ regions as $\gamma^a$, and that between $\mathcal{N}=1$ and $\mathcal{N}=2$ regions as $\gamma^b$. When the system is in the QAHI phase with the average Dirac mass $\langle m\rangle$ sufficiently positive, the $\mathcal{N}=2$ region dominates and is extended as shown in Fig. \ref{domain}(b). As $\langle m\rangle$ decreases into a certain interval in the vicinity of $0$, the system enters a phase where the $\mathcal{N}=1$ region becomes extended, while both the $\mathcal{N}=0$ and $2$ regions are bounded into finite size islands, as shown in Fig. \ref{domain}(c). This is exactly the chiral TSC phase in the presence of disorder. When $\langle m\rangle$ further decreases to sufficiently negative, the $\mathcal{N}=0$ region becomes extended (Fig. \ref{domain}(d)), and the system enters a NI phase. Therefore, in this percolation picture, the system still has three phases $\mathcal{N}=2,1$ and $0$, and undergoes two phase transitions similar to that predicted by the band theory. Accordingly, the chiral Majorana edge states $\gamma^b$ and $\gamma^a$ at zero energy become delocalized exactly at the two phase transition points, respectively.

During the flipping of magnetic domains, one expects the spatial fluctuations in the Dirac mass $m$ to be greater than the fluctuations in $\mu$ and $\Delta$. In this case, the $\mathcal{N}=1$ regions are generically narrower compared with $\mathcal{N}=0$ and $\mathcal{N}=2$ regions, as shown in Fig. \ref{domain}. This may introduce a nonzero hopping between $\gamma^a$ and $\gamma^b$ when they are separated by a narrow $\mathcal{N}=1$ region, which can be characterized by the following effective 1D Hamiltonian:
\begin{equation}\label{hopping}
H_{\text{edge}}=iv_a\gamma^a\partial_l\gamma^a+iv_b\gamma^b\partial_l\gamma^b-2i\eta\gamma^a\gamma^b\ ,
\end{equation}
where $l$ is the 1D coordinate along either of the two nearly parallel neighbouring edges, $v_a>0$ and $v_b>0$ are the velocities of $\gamma^a$ and $\gamma^b$, and $\eta$ is the hopping between $\gamma^a$ and $\gamma^b$ due to their wave function overlap. One may equivalently rewrite the Hamiltonian under the charge basis $\psi=\gamma^a+ i\gamma^b$ and $\psi^\dag=\gamma^a- i\gamma^b$ as
\begin{equation}\label{charge}
H_{\text{edge}}=iv_F\psi^\dag\partial_l\psi-\eta\psi^\dag\psi+(i\Delta_p\psi^\dag\partial_l\psi^\dag+h.c.)\ ,
\end{equation}
where $v_F=(v_a+v_b)/2$, $\Delta_p=(v_a-v_b)/4$ and $\eta$ act as the Fermi velocity, $p$-wave pairing amplitude and chemical potential of the charged chiral fermion $\psi$, respectively. At low energies, the Fermi velocity $v_F=A$, and the $p$-wave pairing is of order $\Delta_p\sim(A/\lambda)\Delta$. When $\Delta_p=0$, $\psi$ simply reduces to the charged chiral edge state of the QAHI. However, the Hamiltonian under charged basis in Eq. (\ref{charge}) does not explicitly reflect the fact that $\gamma^a$ and $\gamma^b$ are \emph{spatially separate} when $\Delta_p\neq0$, which is the key to see there are two phase transitions instead of one. Therefore, we shall work in the Majorana basis $\gamma^a$, $\gamma^b$ hereafter. Below, we will show that the two phase transitions discussed above are unaffected by the inter-edge hopping $2i\eta\gamma^a\gamma^b$ term.

\begin{figure}[tbp]
  \centering
  \includegraphics[width=3.4in]{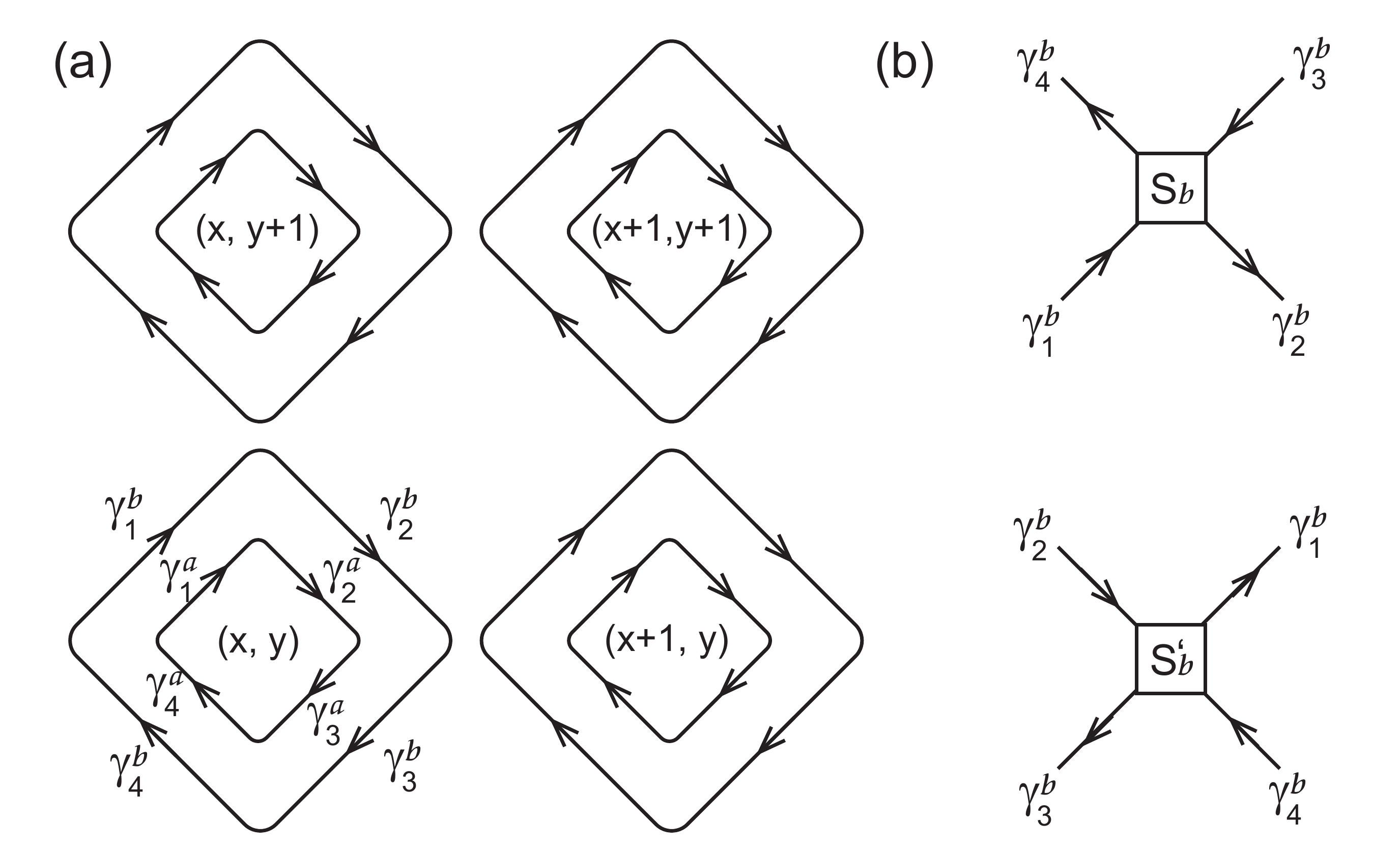}
  \caption{(a) Formulation of the system into a D class network model. (b) Illustration of the scattering of Majorana state $\gamma_b$ at network vertices.}\label{Dnet}
\end{figure}

The percolation of chiral Majorana modes $\gamma^a$ and $\gamma^b$ can be simplified into a 2D network model within the D symmetry class \cite{kramer2005}, which respects the particle-hole symmetry while breaks both the spin rotational and the time reversal symmetries. Fig. \ref{Dnet}(a) shows the network configuration in the vicinity of the phase transition from $\mathcal{N}=2$ to $\mathcal{N}=1$, where the small (large) plaquettes host $N=0$ ($N=1$) regions, and thus $\gamma_a$ ($\gamma_b$) live on the inner (outer) edges surrounding smaller (larger) plaquettes. To the center of each pair of small and large plaquettes we assign a coordinate $(x,y)$, with $x$ and $y$ taking integral values. The four links of a small (large) plaquette are denoted by $i=1,2,3,4$, respectively, and the chiral Majorana edge state on link $i$ is denoted by $\gamma^a_i$ ($\gamma^b_i$). At the vertices in between two large plaquettes, two nearby $\gamma^b$ states at zero energy come close and can scatter into each other. The scattering at vertices located at $(x,y+\frac{1}{2})$ can be described by the scattering matrix
\begin{equation}\label{Sb}
\left(\begin{array}{c}\gamma_2^b\\ \gamma_4^b\end{array}\right)=\left(\begin{array}{cc}\cos\theta_b& \sin\theta_b\\ -\sin\theta_b&\cos\theta_b\end{array}\right) \left(\begin{array}{c}\gamma_1^b\\ \gamma_3^b\end{array}\right)\ ,
\end{equation}
while at vertices located at $(x+\frac{1}{2},y)$
\begin{equation}\label{Sb'}
\left(\begin{array}{c}\gamma_1^b\\ \gamma_3^b\end{array}\right)=\left(\begin{array}{cc}\cos\theta_b& \sin\theta_b\\ -\sin\theta_b&\cos\theta_b\end{array}\right) \left(\begin{array}{c}\gamma_4^b\\ \gamma_2^b\end{array}\right)\ ,
\end{equation}
where the angle $\theta_b\in[0,\pi/2]$ is a function of $(x,y)$. We denote these two matrices as $S_b(\theta_b)$ and $S_b'(\theta_b)$, respectively. Similarly, there are also an angle $\theta_a$ and two matrices $S_a(\theta_a)$, $S_a'(\theta_a)$ of the same form describing the scattering of two nearby $\gamma^a$ states at all vertices. An obvious critical point is $\theta_a=\pi/4$ ($\theta_b=\pi/4$), when $\gamma_a$ ($\gamma_b$) has equal probabilities to scatter in either direction. In the vicinity of the $\pi/4$ critical point, the two scattering angles take the form $\theta_b=\pi/4-\tau_c(m-\sqrt{\mu^2+|\Delta|^2})$ and $\theta_a=\pi/4-\tau_c(m+\sqrt{\mu^2+|\Delta|^2})$ for small $(m\mp\sqrt{\mu^2+|\Delta|^2})$, respectively, which are both linear in the local value of $m$, where $\tau_c>0$ is a constant depending on details of the scattering process. Note that this means $\theta_a\le\theta_b$ in the entire space, so their spatial averages also generically satisfy $\langle\theta_a\rangle<\langle\theta_b\rangle$.

In addition to the scattering matrices at vertices, the network model also has a propagator on each link of a plaquette. In the basis of Majorana fermions $\gamma^a$ and $\gamma^b$, the propagator is real. According to Eq. (\ref{hopping}), the hopping term $\eta$ leads to an amplitude $\gamma^a$ and $\gamma^b$ to scatter into each other, so the propagator on a link, for instance, on link $1$ of plaquette $(x,y)$ at zero energy takes the form
\begin{equation}
\left(\begin{array}{c}\gamma_1^a\\ \gamma_1^b\end{array}\right)_{x-\frac{1}{2},y}= \left(\begin{array}{cc}\cos\phi_1& \sin\phi_1\\ -\sin\phi_1&\cos\phi_1\end{array}\right) \left(\begin{array}{c}\gamma_1^a\\ \gamma_1^b\end{array}\right)_{x,y+\frac{1}{2}},
\end{equation}
where the propagation angle $\phi_1$ depends on $\eta$ and the length $l_1$ of link $1$. Similarly, we can define the propagation angles $\phi_i$ on links $i=2,3,4$. This constitutes a D symmetry class O$(n)$ orthogonal random network model \cite{kramer2005}, where $n$ is the number of flavors of Majorana fermions ($n=2$ here). We note that in general a Majorana fermion $\gamma^a$ could also scatter into $\gamma^b$ at a vertex, or vice versa, but this mutual scattering can be effectively regarded as a propagation of $\gamma^a$ into $\gamma^b$ followed by a scattering from $\gamma^b$ to $\gamma^b$. Therefore, we can simply assume there is no mutual scattering without loss of generality.

\begin{figure}[tbp]
  \centering
  \includegraphics[width=3.4in]{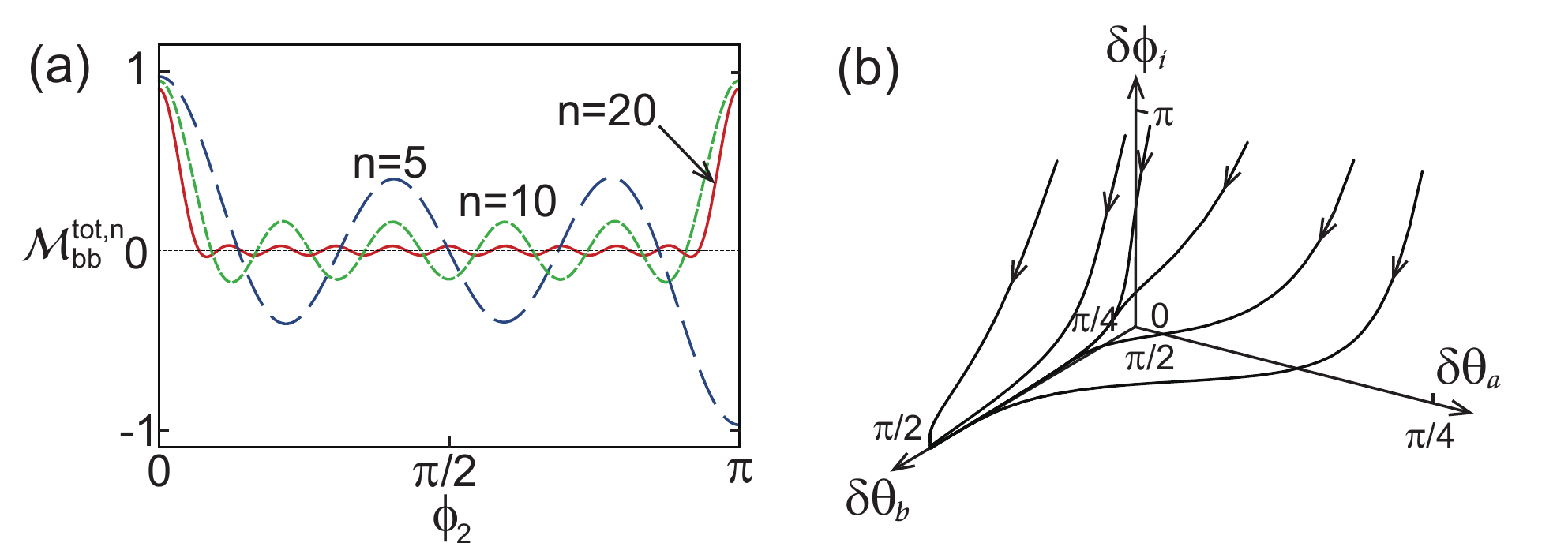}
  \caption{(a) The propagation amplitude $\mathcal{M}_{bb}^{tot,n}$ of $\gamma^b$ near its delocalization transition, which has two peaks at $\phi_2=0$ and $\pi$. (b) Illustration of the RG flow of $\theta_a$, $\theta_b$ and $\phi_i$ near the delocalization transition of $\gamma^b$.}\label{RGflow}
\end{figure}

If all the angles $\phi_i$ are either $0$ or $\pi$, $\gamma^a$ and $\gamma^b$ will never propagate into each other, and the above network model simply decouples into two copies of the minimal O$(1)$ random network model \cite{chalker2001}, which exhibits two delocalization phase transitions at $\langle \theta_a\rangle=\pi/4$ and $\langle \theta_b\rangle=\pi/4$, respectively, while all the states are localized otherwise. If $\phi_i$ are not restricted to $0$ or $\pi$, the two copies of network models are coupled as an O$(2)$ random network model. This, however, does not affect the fact that there are two delocalization phase transitions, as is ensured by the D class symmetry \cite{kramer2005}. Here we present a simplified real space RG analysis to show this fact. Consider the tunneling of Majorana states from vertex $(x-\frac{1}{2},y)$ to vertex $(x+n-\frac{1}{2},y)$, where $n>1$ is an integer. As a great simplification, we ignore the spatial variations of $\theta_a$, $\theta_b$ and $\phi_i$, and assume only $\phi_2$ is nonzero. For convenience, we define $\delta\theta_b=\theta_b-\pi/4$ and $\delta\theta_a=\theta_a-\pi/4$, and assume $|\delta\theta_b|<|\delta\theta_a|$, in which case $\gamma^b$ is more delocalized than $\gamma^a$. The propagation amplitude $\mathcal{M}_{\alpha\beta}^n$ from $\gamma^\beta$ to $\gamma^\alpha$ ($\alpha,\beta=a,b$) contributed by the shortest path is then given by the $2\times2$ matrix
\begin{equation}
\mathcal{M}^n=\left(\begin{array}{cc}\cos\theta_a\sin\theta_a\cos\phi_2& \cos\theta_b\sin\theta_a\sin\phi_2\\ -\cos\theta_a\sin\theta_b\sin\phi_2 &\cos\theta_b\sin\theta_b\cos\phi_2\end{array}\right)^n,
\end{equation}
where the basis is $(\gamma^a, \gamma^b)^T$.
In particular, the amplitude from $\gamma^b$ to $\gamma^b$ reaches its maximum $|\mathcal{M}_{bb}^n|=1/2^n$ when $\theta_b=\pi/4$ and $\phi_2=0$ or $\pi$. On the other hand, we know $\gamma^b$ is delocalized at $\theta_b=\pi/4$ and $\phi_2=0$ or $\pi$, in which case the total propagation amplitude $\mathcal{M}_{bb}^{tot,n}$ should not decay. Therefore, we approximate the total amplitude as $\mathcal{M}_{bb}^{tot,n}\approx 2^n \mathcal{M}_{bb}^n$, which can be thought of as the summation of $2^n$ effective paths similar the shortest path.

When $n$ grows large, the amplitude $\mathcal{M}_{bb}^{tot,n}$ becomes peaked at $\phi_2=0$ and $\pi$ as shown in Fig. \ref{RGflow}(a), with a peak width around $\pi/n$. We then regard $\mathcal{M}_{bb}^{tot,n}$ across $n$ plaquettes as the propagation amplitude across a single renormalized plaquette. Matching the peak value of the amplitude at $\phi=0$ or $\pi$ gives a renormalized $\theta_b'$ satisfying $\sin(2\theta_b')=\sin^n (2\theta_b)$. Furthermore, we define $\delta\phi_2$ to be the relevant range of the propagation angle $\phi_2$ centered at $0$ or $\pi$, which can be identified with the peak width of $\mathcal{M}_{bb}^{tot,n}$, namely, $\delta\phi_2=\pi/n$. Therefore, the only relevant $\phi_2$ at large distances are $0$ and $\pi$, while $\phi_2$ with other values has an amplitude quickly decaying as a function of $n$. For small $\delta\theta_b$ and $\delta\phi_2$, we arrive at the following RG flow equations:
\begin{equation}\label{RGeq}
d\delta\theta_b/dt=\delta\theta_b/2\ ,\quad d\delta\phi_2/dt=-\delta\phi_2\ ,
\end{equation}
where $dt=dn/n$. In general, we expect similar RG flow equations to hold qualitatively for $\theta_a$ and other $\phi_i$. As an example, the RG flow of the system for $\langle\theta_a\rangle> \langle\theta_b\rangle\ge\pi/4$ is as illustrated in Fig. \ref{RGflow}(b). As one can see, at low energies $\phi_i$ flows to $0$ or $\pi$, so the system tends to two decoupled O$(1)$ random network models, and has two phase transitions at $\theta_a=\pi/4$ and $\theta_b=\pi/4$, respectively.

\section{Size and Temperature dependence}\label{Critical}

The localization length $\xi_\alpha$ of Majorana states $\gamma^\alpha$ in the D class network model obeys a critical behavior $\xi_\alpha\propto|\langle\theta_\alpha\rangle-\pi/4|^{-\nu_D}=|\langle \delta\theta_\alpha\rangle|^{-\nu_D}$, with $\alpha=a,b$. In the simplified RG picture above, the RG flow equation (\ref{RGeq}) leads to a critical exponent $\nu_D=2$; while direct numerical calculations of the D class network model show $\nu_D$ is close to $1$ \cite{chalker2001,fulga2012}. In addition, in the vicinity of the critical point $\pi/4$, both $\langle\delta\theta_a\rangle$ and $\langle\delta\theta_b\rangle$ are linear in the average Dirac mass $\langle m\rangle$ and therefore linear in the magnetic field $B$, namely, $\langle\delta\theta_\alpha\rangle\propto(B-B^*_\alpha)$, where $B^*_\alpha$ is the magnetic field where $\gamma^\alpha$ delocalizes. Therefore, the localization length $\xi_\alpha\propto|B-B^*_\alpha|^{-\nu_D}$.

\begin{figure}[tbp]
  \centering
  \includegraphics[width=3.4in]{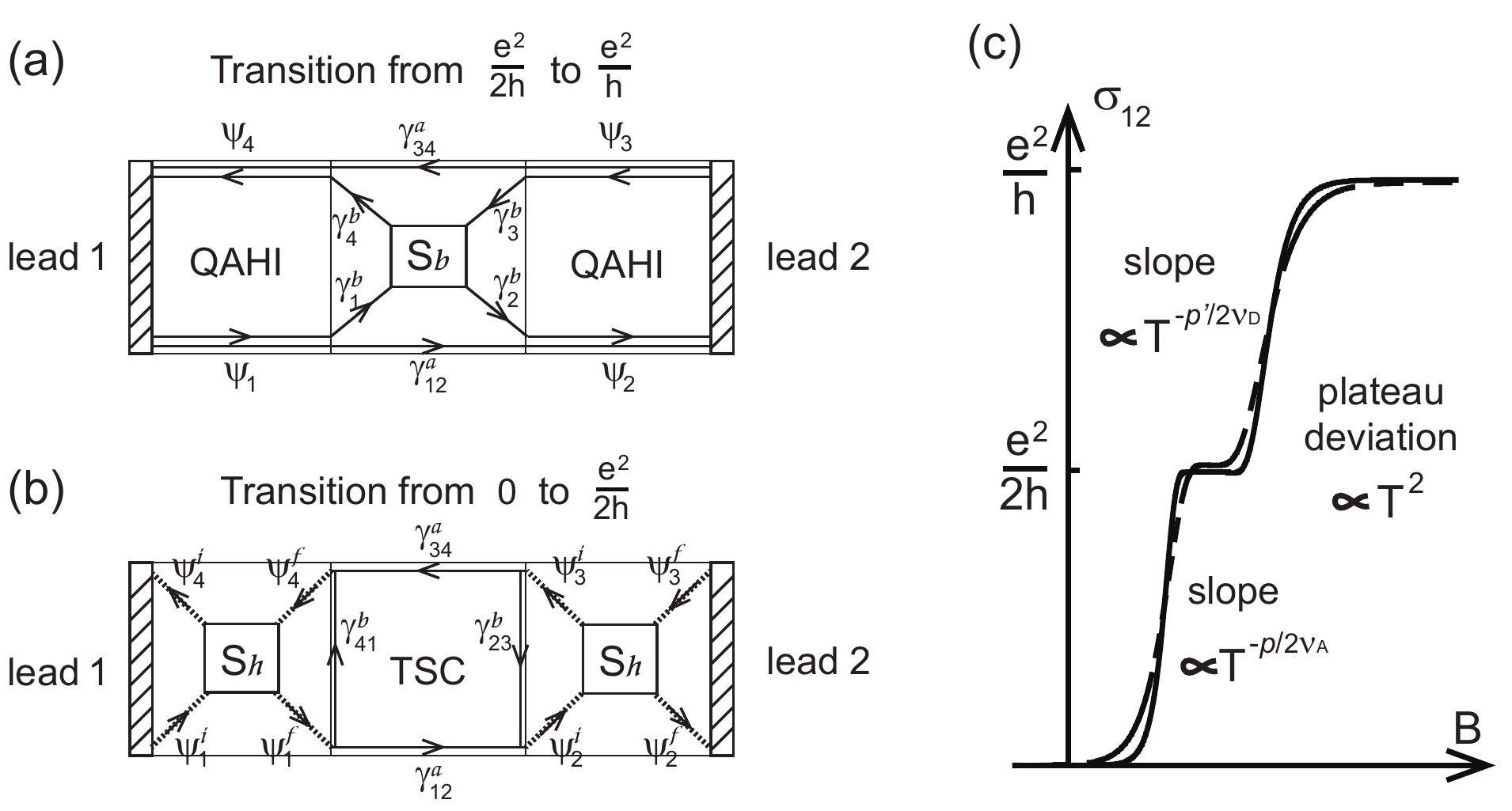}
  \caption{(a) The physical picture of phase transition from $e^2/2h$ to $e^2/h$ after RG flow. (b) The picture of transition from $0$ to $e^2/2h$ after RG flow. (c) Predicted temperature dependence and critical behavior of $\sigma_{12}$ near the half plateau.}\label{Scaling}
\end{figure}

In the vicinity of the plateau transition of $\sigma_{12}$ from $e^2/2h$ to $e^2/h$, $\langle\theta_a\rangle$ is far below $\pi/4$ while $\langle\theta_b\rangle$ is close to $\pi/4$. At zero temperature, as $\langle\theta_b\rangle$ approaches $\pi/4$, the system flows under RG to an effective configuration as shown in Fig. \ref{Scaling}(a), where all the Majorana edge states $\gamma^a$ are localized in the bulk except for the two edge states $\gamma^a_{12}$ and $\gamma_{34}^a$, while the percolation of $\gamma^b$ in the middle region is effectively described by a single renormalized scattering matrix $S_b(\theta^*_b)$ as
\begin{equation}
\left(\begin{array}{c}\gamma_2^b\\ \gamma_4^b\end{array}\right)=\left(\begin{array}{cc}\cos\theta_b^*& \sin\theta_b^*\\ -\sin\theta_b^*&\cos\theta_b^*\end{array}\right) \left(\begin{array}{c}\gamma_1^b\\ \gamma_3^b\end{array}\right)\ ,
\end{equation}
where $\theta_b^*\in[0,\pi/2]$ is the renormalized scattering angle of $\gamma^b$ at length scale of the size of the TSC region $L$. The scaling behavior of $\delta\theta_b^*=\theta_b^*-\pi/4$ suggests it is an analytical function $\delta\theta_b^*=f_b(\kappa_b)$ of the dimensionless parameter $\kappa_b=[L/\xi_b(B)]^{1/\nu_D}$, where $\xi_b(B)$ is the localization length of $\gamma^b$ at magnetic field $B$ \cite{pruisken1984,pruisken1988}. For small $\kappa_b$, $\delta\theta_b^*$ can be expanded into
\begin{equation}
\delta\theta_b^*=f_b(\kappa_b)=b_1\kappa_b+b_2\kappa_b^2+\cdots.
\end{equation}
Note that $\kappa_b\propto L^{-\nu_D}(B-B^*_b)$ at small $\kappa_b$, so this agrees with the fact that $\delta\theta^*_b\propto\langle\delta\theta_b\rangle\propto(B-B^*_b)$.

The two terminal conductance $\sigma_{12}$ can be expressed as a function of $\theta_b^*$. The charged fermion operators in Fig. \ref{Scaling}(a) are related to the Majorana operators as $\psi_1=\gamma^a_{12}+i\gamma_1^b$, $\psi_2=\gamma^a_{12}+i\gamma_2^b$, $\psi_3=\gamma^a_{34}+i\gamma_3^b$ and $\psi_4=\gamma^a_{34}+i\gamma_4^b$. Therefore, the normal (Andreev) transmission (reflection) probabilities of the junction are given by $t=|\langle\psi_2^\dag\psi_1 \rangle|^2=(1+\cos\theta^*_b)^2/4$, $t_A=|\langle\psi_2\psi_1 \rangle|^2=(1-\cos\theta^*_b)^2/4$, $r=|\langle\psi_4^\dag\psi_1 \rangle|^2=\sin^2\theta^*_b/4$ and $r_A=|\langle\psi_4\psi_1 \rangle|^2=\sin^2\theta^*_b/4$, respectively. The conductance $\sigma_{12}=(t+r_A)e^2/h$ according to the generalized Landauer-B\"uttiker formula \cite{anantram1996,entin2008} is then
\begin{equation}
\sigma_{12}=\frac{1+\cos\theta_b^*}{2}\frac{e^2}{h}\ .
\end{equation}
In particular, one finds the slope of $\sigma_{12}$ as a function of $B$ near the critical point $\theta_b^*=\pi/4$ has a critical scaling behavior
\begin{equation}
\frac{d\sigma_{12}}{dB}\propto \frac{d\delta\theta_b^*}{dB} \propto L^{1/\nu_D}\ ,
\end{equation}
in analogy to the scaling behavior of the integer quantum Hall effect (IQHE) plateau transition \cite{pruisken1988}. In the regime $L>\xi_b$, namely $\kappa_b>1$, the scattering amplitude between $\gamma_b$ on opposite edges is expected to be proportional to $e^{-L/\xi_b}$, and one expects $\sigma_{12}$ to behave as \cite{pruisken1988}
\begin{equation}
|\sigma_{12}-\sigma_{12}^0|\propto e^{-|\kappa_b|^{\nu_D}}=e^{-L/\xi_b}\ ,
\end{equation}
where $\sigma_{12}^0$ is either $e^2/2h$ or $e^2/h$ depending on which side of plateau transition the system is on.

The plateau transition from $e^2/2h$ to $0$, however, is in a different universality class, which happens at $\langle m\rangle\propto(B-B_0)=0$. This transition is induced by the phase transition from QAHI to NI, namely Chern number $\mathcal{C}=1$ to $\mathcal{C}=0$, in the left and right regions of the junction, which is in the A class \cite{kramer2005}, while the middle region remains deep in the $\mathcal{N}=1$ chiral TSC phase. The QAHI to NI transition is governed by an A symmetry class network model describing the delocalization transition of charged chiral states \cite{chalker1988}, where a scattering angle $\theta_h$ and scattering matrix $S_h(\theta_h)$ can be defined at each vertices of the network similar to Eq. (\ref{Sb}). The only difference of the A class network model from a D class network model is its propagation angles $\phi_i$ are arbitrary instead of being only $0$ or $\pi$, while the delocalization transition is still at $\theta_h=\pi/4$. When $\theta_h$ deviates from $\pi/4$, the localization length $\xi_h$ of the charged chiral states obeys the critical behavior $\xi_h\propto |\langle\delta \theta_h\rangle|^{-\nu_A}$, where $\delta \theta_h=\theta_h-\pi/4$, and $\nu_A\approx7/3$ as revealed by numerical calculations \cite{chalker1988}. Physically, $\delta\theta_h$ is proportional to the Dirac mass $m$, and thus its average value $\langle\delta \theta_h\rangle\propto\langle m\rangle\propto B-B_0$.

Similarly, under RG the system in the vicinity of the QAHI to NI transition flows to a configuration as shown in Fig. \ref{Scaling}(b), where both the left and right regions are described by an effective scattering matrix
\begin{equation}
S_h(\theta^*_h)=\left(\begin{array}{cc}\cos\theta_h^*& \sin\theta_h^*\\ -\sin\theta_h^*&\cos\theta_h^*\end{array}\right)\ ,
\end{equation}
with $(\psi^f_4,\psi^f_1)^T=S_h(\theta^*_h)(\psi^i_4,\psi^i_1)^T$, and $(\psi^f_3,\psi^f_2)^T=S_h(\theta^*_h)(\psi^i_3,\psi^i_2)^T$. The renormalized scattering angle $\theta^*_h=f_h(\kappa_h)$ is an analytical function of dimensionless parameter $\kappa_h=[L/\xi_h(B)]^{1/\nu_A}\propto L^{1/\nu_A}(B-B_0)$, where $L$ being the size of the QAHI regions. Besides, each charged chiral state has a random propagation phase. The normal (Andreev) transmission (reflection) probabilities $t=|\langle\psi_2^{f\dag}\psi_1^i \rangle|^2$, $t_A=|\langle\psi_2^{f}\psi_1^i \rangle|^2$, $r=|\langle\psi_4^{f\dag}\psi_1^i \rangle|^2$ and $r_A=|\langle\psi_2^{f}\psi_1^i \rangle|^2$ can be calculated by summing over all the propagation probabilities via various paths, which is rather difficult. However, the two terminal conductance $\sigma_{12}=(t+r_A)e^2/h$ can be calculated much more easily by noting that each path containing chiral Majorana edge states of the TSC contributes equally to $t$ and $t_A$, and also equally to $r$ and $r_A$. For an incident electron from state $\psi_1^i$, the only path not involving chiral Majorana edge states is the shortest path from $\psi_1^i$ to $\psi_4^f$, which contributes a probability $\sin^2\theta_h^*$ only to $r$. Therefore one has $t=t_A$, and $r-r_A=\sin^2\theta_h^*$. Besides, the unitarity puts a restriction $t+t_A+r+r_A=1$, from which one finds
\begin{equation}
\sigma_{12}=(t+r_A)\frac{e^2}{h}=\frac{\cos^2\theta_h^*}{2}\frac{e^2}{h}\ .
\end{equation}
In the vicinity of the A class critical point $\theta_h^*=\pi/4$, an expansion with respect to $\delta\theta_h^*=\theta_h^*-\pi/4$ yields
\begin{equation}
\frac{d\sigma_{12}}{dB}\propto \frac{d\delta\theta_h^*}{dB}\propto L^{1/\nu_A}\ .
\end{equation}
While when the system is far away from the critical point, one has \cite{pruisken1988}
\begin{equation}
|\sigma_{12}-\sigma_{12}^0|\propto e^{-|\kappa_h|^{\nu_A}}=e^{-L/\xi_h}\ ,
\end{equation}
where $\sigma_{12}^0$ is either $e^2/2h$ or $0$.

%\cite{biscaras2013,sondhi1997}

At finite temperatures $T$, the size $L$ of the junction should be replaced by when the inelastic scattering length \cite{thouless1977} $L_{in}$ of the QAHI or $L_{in}'$ of the TSC is smaller than the system size $L$, one should replace $L$ by $L_{in}$ and $L_{in}'$ in the plateau transition from $e^2/2h$ to $0$ and that from $e^2/2h$ to $e^2/h$, respectively. Due to the scale invariance at quantum phase transitions, the inelastic scattering lengths generically obey power law scaling $L_{in}\propto T^{-p/2}$ and $L_{in}'\propto T^{-p'/2}$, where $p>0$ and $p'>0$ are critical exponents of inelastic scattering for A class and D class systems, respectively \cite{sondhi1997}. In IQHE or QAHI systems, it is known $p=2$ in the homogeneous limit, and $p=1$ in the disordered limit \cite{pruisken1988,huckestein1990,kou2015}. For 2D superconductors, the critical exponent is approximately $p'\approx 2$ (more often denoted as $z=(p'/2)^{-1}\approx1$) \cite{sondhi1997,biscaras2013}. Assuming the slope of exchange field $\langle\lambda\rangle$ with respect to $B$ approaches a constant as $T\rightarrow0$ (as is true for ferromagnetic materials), we arrive at the temperature dependence of the slope of two terminal conductance $\sigma_{12}$
\begin{equation}
\frac{d\sigma_{12}}{dB}\propto L_{in}^{1/\nu_A} \propto T^{-p/2\nu_A}
\end{equation}
from $0$ to $e^2/2h$, and
\begin{equation}
\frac{d\sigma_{12}}{dB}\propto L_{in}'^{1/\nu_D}\propto T^{-p'/2\nu_D}
\end{equation}
from $e^2/2h$ to $e^2/h$.

In addition, a finite temperature $T$ also drives the half conductance plateau away from the quantized value $\sigma_{12}=e^2/2h$. This is because the conductance is contributed by quasi-particles within a finite energy window $k_BT$ near the fermi surface, while the chiral Majorana edge modes at finite energies are no longer charge neutral. When an electron $\psi_{i,\epsilon}$ ($1\le i\le 4$) at finite energy $\epsilon$ encounters the boundary of the TSC as shown in Fig. \ref{domain}(a), it splits into $\gamma_{i,\epsilon}$ and $\gamma_{i+1,\epsilon}$ in the way $\psi_{i,\epsilon}=u_{i}(\epsilon)\gamma_{i,\epsilon}+iv_{i}(\epsilon)\gamma_{i+1,\epsilon}$ (label $i$ is identified with $i+4$), where the coefficients satisfy $u_{i}(0)=v_{i}(0)=1$. The transmission and Andreev reflection probability from $\psi_{1,\epsilon}$ to $\psi_{2,\epsilon}$ and to $\psi_{4,\epsilon}$ are given by $t(\epsilon)=|v_1^*(\epsilon)u_2(\epsilon)|^2/4$ and $r_A(\epsilon)=|u_1(\epsilon)v_4(\epsilon)|^2/4$, respectively. The two terminal conductance $\sigma_{12}$ of the junction is then given by
\begin{equation}
\sigma_{12}=\int d\epsilon \left[-\frac{df(\epsilon)}{d\epsilon}\right]\left[t(\epsilon)+r_A(\epsilon)\right]\ ,
\end{equation}
where $f(\epsilon)$ is the Fermi Dirac distribution at temperature $T$. For small $\epsilon$ one has $\left[t(\epsilon)+r_A(\epsilon)\right]=1/2+c_1\epsilon+c_2\epsilon^2$, and an easy calculation gives $\delta\sigma_{12}=\sigma_{12}-e^2/2h\propto T^2$.

Recently, two papers \cite{ji2017,huang2017} have discussed potential alternative explanations to the half plateau of $\sigma_{12}$ observed in the experiment \cite{he2017}.
In both papers, the middle region of the junction is formulated as a wiggling 1D charged chiral state based on the classical percolating picture, with random pairing amplitude $\Delta(\mathbf{r})$ perturbations added, and the critical behavior of the A class is used in the work of Huang \emph{et al.}\cite{huang2017}. They conclude a half plateau arises when the effective length of the 1D state is much longer than the system size. We first emphasize that the system should always be treated within the D class as long as the pairing $\Delta(\mathbf{r})$ is nonzero. In particular, the two chiral Majorana states $\gamma^a$ and $\gamma^b$ are spatially separated under the RG flow, therefore cannot be regarded as a single charged chiral state. Accordingly, the single A class phase transition from Chern number $\mathcal{C}=1$ to $0$ is always unstable when $\Delta(\mathbf{r})$ is nonzero, and will split into two phase transitions from BdG Chern number $\mathcal{N}=2$ to $1$ and from $\mathcal{N}=1$ to $0$. Secondly, the picture of a single wiggling 1D chiral state assumed in both papers may not be able to capture the exact physics of this system, which is intrinsically 2D. The quantum tunnelings between spatially nearby parts of the wiggling path of the chiral state are ignored in this picture, which may significantly affect the critical behavior. Further, their theories lead to several predictions different from ours as we shall illustrate below, which can be tested in the experiments.

In the work of Ji and Wen \cite{ji2017}, they argue the dephasing together with the finite system size under strong disorders can also lead to a half conductance plateau, which vanishes as the system size $L\rightarrow\infty$, and there is only one delocalization phase transition directly from BdG Chern number $\mathcal{N}=0$ to $\mathcal{N}=2$. They predict the half plateau occurs when the classical localization length $\xi$ in the QAHI is larger than $L$. The scaling law of the classical localization length is given by $\xi\propto|B-B_0|^{-\nu}$ with $\nu=4/3$. This leads to a plateau width $\Delta B_{1/2}\propto L^{-1/\nu}$. This is a key difference from our D class theory, where we predict an intermediate $\mathcal{N}=1$ TSC phase and a finite half plateau width as $L\rightarrow\infty$. For the experiment \cite{he2017}, $L=1$mm, their theory would predict a plateau width way too small. The domain wall measurement of Cr doped (Bi,Sb)$_2$Te$_3$ thin films \cite{lachmane2015} reveals that the magnetic domain size is of order $0.1\mu$m at a magnetic field $50$mT away from the percolation transition. If we identify the domain size as $\xi$ at this magnetic field, their theory yields a plateau width %\cite{nuA}
$\Delta B_{1/2}\sim (50$mT$)(L/0.1\mu$m$)^{-1/\nu}\lesssim0.3$mT, while the observed plateau width is much larger around $20$mT. We expect future experiments on the size dependence of half plateau width to further confirm our theory.

Besides, their theory implicitly predicts the slope of plateau transition $d\sigma_{12}/dB$ from $0$ to $e^2/2h$ and from $e^2/2h$ to $e^2/h$ depends on $(V^2/l_{\phi})$ when the temperature $T<eV/k_B$, where $l_\phi$ is the dephasing length of the system, and $V$ is the voltage applied between leads $1$ and $2$ for measuring $\sigma_{12}$.
%Since $l_\phi$ increases as temperature $T$ lowers, this leads to a slower slope at lower temperatures, which is just \emph{opposite} to our prediction.
In contrast, the slope $d\sigma_{12}/dB$ near plateau transitions does not have a voltage dependence in our theory; instead it only depends on the system size and temperature.

In another work of Huang \emph{et al.} \cite{huang2017}, they support our conclusions that an intermediate $\mathcal{N}=1$ TSC phase should exist in between the $\mathcal{N}=0$ and $\mathcal{N}=2$ phases, while they argue $\sigma_{12}$ could be well quantized to the half plateau way before the system enters the $\mathcal{N}=1$ phase. At the plateau transitions from $e^2/2h$ to $0$ or $e^2/h$, they obtain a critical behavior very different from ours, which reads $\sigma_{12}=\left(1\pm e^{-(l_h/l_0)}\right)e^2/2h$, where $l_h\propto\xi_h^{d_f}$ is the circumference of a 1D cluster of size $\xi_h$, $\xi_h$ is the QAHI localization length we discussed ahead, $d_f=91/48$ is the fractal dimension of 2D clusters, and $l_0$ is some characteristic length of the system. The slope of such a plateau transition is expected to have a different dependence on temperature $T$ and system size $L$ from that in our theory.

In the end of their paper they also mentioned another possibility that the intermediate phase could be a thermal metal instead of an $\mathcal{N}=1$ TSC under strong disorders \cite{senthil2000,chalker2001,medvedyeva2010,wimmer2010,fulga2012}, which also leads to a half plateau. However, numerical studies show such a thermal metal phase is possible only if there are sufficient localized random Majorana zero modes in the D class network model, which can be induced either by $\pi$ flux superconducting vortices \cite{chalker2001} or a strong electrostatic disorder potential \cite{medvedyeva2010,wimmer2010,fulga2012}. In particular, the electrostatic disorder potential has to be comparable to the spin-orbit coupling (SOC) strength or the conduction (valence) band width of the system \cite{wimmer2010}, which is usually much larger than the bulk gap. In the realistic system, the creation of $\pi$ flux vortices will be energetically unfavorable since the $s$-wave SC on top of the QAHI is thick ($\sim200$nm). Besides, although the bulk gap of the Cr-doped (Bi,Sb)$_2$Te$_3$ system under superconducting proximity is small, its SOC strength and band width is of the order of $1$ eV \cite{liu2010}, which is expected to be much larger than the electrostatic disorder potential in the system. Therefore, we conclude the thermal metal phase is extremely unlikely to occur in the experiment \cite{he2017}.

\section{Conclusion}\label{Conclusion}
In this paper, we studied the phase transitions of disordered QAHI system in proximity of an $s$-wave superconductor employing the percolation theory of chiral Majorana fermions, which can be written as a random network model in the D symmetry class. Based on symmetry arguments and a simple RG flow analysis, we show there are generically two phase transitions separating three superconducting phases carrying BdG Chern numbers $\mathcal{N}=2,1$ and $0$, respectively, which agrees with the conclusion from the simple band theory of homogeneous systems \cite{qi2010b}. In particular, the $\mathcal{N}=1$ phase, namely, the $p+ip$ chiral TSC phase, is robust against disorders.

%In practice, the two phase transitions are driven by the spatial average value of the Dirac mass $\langle m\rangle$, which is linear to the magnetic field $B$.

For a QAHI bar in proximity with an $s$-wave SC in its middle region (Fig. \ref{domain}(a)), it has been shown in the absence of disorders that the two terminal conductance $\sigma_{12}$ exhibits a half plateau at $e^2/2h$ in between two integer plateaus at $0$ and $e^2/h$ as a function of $B$, which indicates the formation of a $p+ip$ chiral TSC \cite{chung2011,wang2015a}. Since we have shown the $p+ip$ chiral TSC phase is stable against disorders, we conclude the half plateau picture still holds in inhomogeneous systems, which explains the half plateau recently observed in disordered Cr doped (Bi,Sb)$_2$Te$_3$ QAHI thin films under SC proximity \cite{he2017}. We then study the critical scaling behavior of the half plateau and the plateau transitions governed by the percolation theory with respect to system size and temperature. In particular, the plateau transition from $0$ to $e^2/2h$ and that from $e^2/2h$ to $e^2/h$ belong to the A and D symmetry classes, respectively, and have different critical exponents. Finally, we discussed the differences between our results and two recent alternative explanations of half plateau \cite{ji2017,huang2017}, and showed the critical behaviors we predicted are quite distinct from those predicted in their theories, which are testable in current and future experiments.

\

\begin{acknowledgments}
We thank C. Beenakker for his helpful communications with us. B.L. is supported by Princeton Center for Theoretical Science at Princeton University. J.W. is supported by the Natural Science Foundation of China through Grant No.~11774065 and by the Natural Science Foundation of Shanghai under Grant No.~17ZR1442500. A.V. was funded by the Gordon and Betty Moore Foundation's EPiQS Initiative through Grant GBMF4302. S.C.Z. is supported by the NSF under grant numbers DMR-1305677
and the US Department of Energy, Office of Basic Energy Sciences
under contract DE-AC02-76SF00515.
\end{acknowledgments}

%\bibliography{CM_ref}

\begin{thebibliography}{50}%
\makeatletter
\providecommand \@ifxundefined [1]{%
 \@ifx{#1\undefined}
}%
\providecommand \@ifnum [1]{%
 \ifnum #1\expandafter \@firstoftwo
 \else \expandafter \@secondoftwo
 \fi
}%
\providecommand \@ifx [1]{%
 \ifx #1\expandafter \@firstoftwo
 \else \expandafter \@secondoftwo
 \fi
}%
\providecommand \natexlab [1]{#1}%
\providecommand \enquote  [1]{``#1''}%
\providecommand \bibnamefont  [1]{#1}%
\providecommand \bibfnamefont [1]{#1}%
\providecommand \citenamefont [1]{#1}%
\providecommand \href@noop [0]{\@secondoftwo}%
\providecommand \href [0]{\begingroup \@sanitize@url \@href}%
\providecommand \@href[1]{\@@startlink{#1}\@@href}%
\providecommand \@@href[1]{\endgroup#1\@@endlink}%
\providecommand \@sanitize@url [0]{\catcode `\\12\catcode `\$12\catcode
  `\&12\catcode `\#12\catcode `\^12\catcode `\_12\catcode `\%12\relax}%
\providecommand \@@startlink[1]{}%
\providecommand \@@endlink[0]{}%
\providecommand \url  [0]{\begingroup\@sanitize@url \@url }%
\providecommand \@url [1]{\endgroup\@href {#1}{\urlprefix }}%
\providecommand \urlprefix  [0]{URL }%
\providecommand \Eprint [0]{\href }%
\providecommand \doibase [0]{http://dx.doi.org/}%
\providecommand \selectlanguage [0]{\@gobble}%
\providecommand \bibinfo  [0]{\@secondoftwo}%
\providecommand \bibfield  [0]{\@secondoftwo}%
\providecommand \translation [1]{[#1]}%
\providecommand \BibitemOpen [0]{}%
\providecommand \bibitemStop [0]{}%
\providecommand \bibitemNoStop [0]{.\EOS\space}%
\providecommand \EOS [0]{\spacefactor3000\relax}%
\providecommand \BibitemShut  [1]{\csname bibitem#1\endcsname}%
\let\auto@bib@innerbib\@empty
%</preamble>
\bibitem [{\citenamefont {Fu}\ and\ \citenamefont {Kane}(2008)}]{fu2008}%
  \BibitemOpen
  \bibfield  {author} {\bibinfo {author} {\bibfnamefont {L.}~\bibnamefont
  {Fu}}\ and\ \bibinfo {author} {\bibfnamefont {C.~L.}\ \bibnamefont {Kane}},\
  }\href {\doibase 10.1103/PhysRevLett.100.096407} {\bibfield  {journal}
  {\bibinfo  {journal} {Phys. Rev. Lett.}\ }\textbf {\bibinfo {volume} {100}},\
  \bibinfo {pages} {096407} (\bibinfo {year} {2008})}\BibitemShut {NoStop}%
\bibitem [{\citenamefont {Fu}\ and\ \citenamefont {Kane}(2009)}]{fu2009a}%
  \BibitemOpen
  \bibfield  {author} {\bibinfo {author} {\bibfnamefont {L.}~\bibnamefont
  {Fu}}\ and\ \bibinfo {author} {\bibfnamefont {C.~L.}\ \bibnamefont {Kane}},\
  }\href {\doibase 10.1103/PhysRevLett.102.216403} {\bibfield  {journal}
  {\bibinfo  {journal} {Phys. Rev. Lett.}\ }\textbf {\bibinfo {volume} {102}},\
  \bibinfo {pages} {216403} (\bibinfo {year} {2009})}\BibitemShut {NoStop}%
\bibitem [{\citenamefont {Sato}\ and\ \citenamefont
  {Fujimoto}(2009)}]{sato2009a}%
  \BibitemOpen
  \bibfield  {author} {\bibinfo {author} {\bibfnamefont {M.}~\bibnamefont
  {Sato}}\ and\ \bibinfo {author} {\bibfnamefont {S.}~\bibnamefont
  {Fujimoto}},\ }\href {\doibase 10.1103/PhysRevB.79.094504} {\bibfield
  {journal} {\bibinfo  {journal} {Phys. Rev. B}\ }\textbf {\bibinfo {volume}
  {79}},\ \bibinfo {pages} {094504} (\bibinfo {year} {2009})}\BibitemShut
  {NoStop}%
\bibitem [{\citenamefont {Sau}\ \emph {et~al.}(2010)\citenamefont {Sau},
  \citenamefont {Lutchyn}, \citenamefont {Tewari},\ and\ \citenamefont
  {Das~Sarma}}]{sau2010}%
  \BibitemOpen
  \bibfield  {author} {\bibinfo {author} {\bibfnamefont {J.~D.}\ \bibnamefont
  {Sau}}, \bibinfo {author} {\bibfnamefont {R.~M.}\ \bibnamefont {Lutchyn}},
  \bibinfo {author} {\bibfnamefont {S.}~\bibnamefont {Tewari}}, \ and\ \bibinfo
  {author} {\bibfnamefont {S.}~\bibnamefont {Das~Sarma}},\ }\href {\doibase
  10.1103/PhysRevLett.104.040502} {\bibfield  {journal} {\bibinfo  {journal}
  {Phys. Rev. Lett.}\ }\textbf {\bibinfo {volume} {104}},\ \bibinfo {pages}
  {040502} (\bibinfo {year} {2010})}\BibitemShut {NoStop}%
\bibitem [{\citenamefont {Alicea}(2010)}]{alicea2010}%
  \BibitemOpen
  \bibfield  {author} {\bibinfo {author} {\bibfnamefont {J.}~\bibnamefont
  {Alicea}},\ }\href {\doibase 10.1103/PhysRevB.81.125318} {\bibfield
  {journal} {\bibinfo  {journal} {Phys. Rev. B}\ }\textbf {\bibinfo {volume}
  {81}},\ \bibinfo {pages} {125318} (\bibinfo {year} {2010})}\BibitemShut
  {NoStop}%
\bibitem [{\citenamefont {Raghu}\ \emph {et~al.}(2010)\citenamefont {Raghu},
  \citenamefont {Kapitulnik},\ and\ \citenamefont {Kivelson}}]{raghu2010}%
  \BibitemOpen
  \bibfield  {author} {\bibinfo {author} {\bibfnamefont {S.}~\bibnamefont
  {Raghu}}, \bibinfo {author} {\bibfnamefont {A.}~\bibnamefont {Kapitulnik}}, \
  and\ \bibinfo {author} {\bibfnamefont {S.~A.}\ \bibnamefont {Kivelson}},\
  }\href {\doibase 10.1103/PhysRevLett.105.136401} {\bibfield  {journal}
  {\bibinfo  {journal} {Phys. Rev. Lett.}\ }\textbf {\bibinfo {volume} {105}},\
  \bibinfo {pages} {136401} (\bibinfo {year} {2010})}\BibitemShut {NoStop}%
\bibitem [{\citenamefont {Tanaka}\ \emph {et~al.}(2009)\citenamefont {Tanaka},
  \citenamefont {Yokoyama},\ and\ \citenamefont {Nagaosa}}]{tanaka2009a}%
  \BibitemOpen
  \bibfield  {author} {\bibinfo {author} {\bibfnamefont {Y.}~\bibnamefont
  {Tanaka}}, \bibinfo {author} {\bibfnamefont {T.}~\bibnamefont {Yokoyama}}, \
  and\ \bibinfo {author} {\bibfnamefont {N.}~\bibnamefont {Nagaosa}},\ }\href
  {\doibase 10.1103/PhysRevLett.103.107002} {\bibfield  {journal} {\bibinfo
  {journal} {Phys. Rev. Lett.}\ }\textbf {\bibinfo {volume} {103}},\ \bibinfo
  {pages} {107002} (\bibinfo {year} {2009})}\BibitemShut {NoStop}%
\bibitem [{\citenamefont {Akhmerov}\ \emph {et~al.}(2009)\citenamefont
  {Akhmerov}, \citenamefont {Nilsson},\ and\ \citenamefont
  {Beenakker}}]{akhmerov2009}%
  \BibitemOpen
  \bibfield  {author} {\bibinfo {author} {\bibfnamefont {A.~R.}\ \bibnamefont
  {Akhmerov}}, \bibinfo {author} {\bibfnamefont {J.}~\bibnamefont {Nilsson}}, \
  and\ \bibinfo {author} {\bibfnamefont {C.~W.~J.}\ \bibnamefont {Beenakker}},\
  }\href {\doibase 10.1103/PhysRevLett.102.216404} {\bibfield  {journal}
  {\bibinfo  {journal} {Phys. Rev. Lett.}\ }\textbf {\bibinfo {volume} {102}},\
  \bibinfo {pages} {216404} (\bibinfo {year} {2009})}\BibitemShut {NoStop}%
\bibitem [{\citenamefont {Mackenzie}\ and\ \citenamefont
  {Maeno}(2003)}]{mackenzie2003}%
  \BibitemOpen
  \bibfield  {author} {\bibinfo {author} {\bibfnamefont {A.~P.}\ \bibnamefont
  {Mackenzie}}\ and\ \bibinfo {author} {\bibfnamefont {Y.}~\bibnamefont
  {Maeno}},\ }\href {\doibase 10.1103/RevModPhys.75.657} {\bibfield  {journal}
  {\bibinfo  {journal} {Rev. Mod. Phys.}\ }\textbf {\bibinfo {volume} {75}},\
  \bibinfo {pages} {657} (\bibinfo {year} {2003})}\BibitemShut {NoStop}%
\bibitem [{\citenamefont {Lutchyn}\ \emph {et~al.}(2010)\citenamefont
  {Lutchyn}, \citenamefont {Sau},\ and\ \citenamefont
  {Das~Sarma}}]{lutchyn2010}%
  \BibitemOpen
  \bibfield  {author} {\bibinfo {author} {\bibfnamefont {R.~M.}\ \bibnamefont
  {Lutchyn}}, \bibinfo {author} {\bibfnamefont {J.~D.}\ \bibnamefont {Sau}}, \
  and\ \bibinfo {author} {\bibfnamefont {S.}~\bibnamefont {Das~Sarma}},\ }\href
  {\doibase 10.1103/PhysRevLett.105.077001} {\bibfield  {journal} {\bibinfo
  {journal} {Phys. Rev. Lett.}\ }\textbf {\bibinfo {volume} {105}},\ \bibinfo
  {pages} {077001} (\bibinfo {year} {2010})}\BibitemShut {NoStop}%
\bibitem [{\citenamefont {Qi}\ and\ \citenamefont {Zhang}(2011)}]{qi2011}%
  \BibitemOpen
  \bibfield  {author} {\bibinfo {author} {\bibfnamefont {X.-L.}\ \bibnamefont
  {Qi}}\ and\ \bibinfo {author} {\bibfnamefont {S.-C.}\ \bibnamefont {Zhang}},\
  }\href {\doibase 10.1103/RevModPhys.83.1057} {\bibfield  {journal} {\bibinfo
  {journal} {Rev. Mod. Phys.}\ }\textbf {\bibinfo {volume} {83}},\ \bibinfo
  {pages} {1057} (\bibinfo {year} {2011})}\BibitemShut {NoStop}%
\bibitem [{\citenamefont {Wang}\ \emph {et~al.}(2013)\citenamefont {Wang},
  \citenamefont {Platt}, \citenamefont {Yang}, \citenamefont {Honerkamp},
  \citenamefont {Zhang}, \citenamefont {Hanke}, \citenamefont {Rice},\ and\
  \citenamefont {Thomale}}]{wangqh2013}%
  \BibitemOpen
  \bibfield  {author} {\bibinfo {author} {\bibfnamefont {Q.~H.}\ \bibnamefont
  {Wang}}, \bibinfo {author} {\bibfnamefont {C.}~\bibnamefont {Platt}},
  \bibinfo {author} {\bibfnamefont {Y.}~\bibnamefont {Yang}}, \bibinfo {author}
  {\bibfnamefont {C.}~\bibnamefont {Honerkamp}}, \bibinfo {author}
  {\bibfnamefont {F.~C.}\ \bibnamefont {Zhang}}, \bibinfo {author}
  {\bibfnamefont {W.}~\bibnamefont {Hanke}}, \bibinfo {author} {\bibfnamefont
  {T.~M.}\ \bibnamefont {Rice}}, \ and\ \bibinfo {author} {\bibfnamefont
  {R.}~\bibnamefont {Thomale}},\ }\href@noop {} {\bibfield  {journal} {\bibinfo
   {journal} {Europhys. Lett.}\ }\textbf {\bibinfo {volume} {104}},\ \bibinfo
  {pages} {17013} (\bibinfo {year} {2013})}\BibitemShut {NoStop}%
\bibitem [{\citenamefont {Law}\ \emph {et~al.}(2009)\citenamefont {Law},
  \citenamefont {Lee},\ and\ \citenamefont {Ng}}]{law2009}%
  \BibitemOpen
  \bibfield  {author} {\bibinfo {author} {\bibfnamefont {K.~T.}\ \bibnamefont
  {Law}}, \bibinfo {author} {\bibfnamefont {P.~A.}\ \bibnamefont {Lee}}, \ and\
  \bibinfo {author} {\bibfnamefont {T.~K.}\ \bibnamefont {Ng}},\ }\href
  {\doibase 10.1103/PhysRevLett.103.237001} {\bibfield  {journal} {\bibinfo
  {journal} {Phys. Rev. Lett.}\ }\textbf {\bibinfo {volume} {103}},\ \bibinfo
  {pages} {237001} (\bibinfo {year} {2009})}\BibitemShut {NoStop}%
\bibitem [{\citenamefont {He}\ \emph {et~al.}(2014)\citenamefont {He},
  \citenamefont {Wu}, \citenamefont {Choy}, \citenamefont {Liu}, \citenamefont
  {Tanaka},\ and\ \citenamefont {Law}}]{he2014}%
  \BibitemOpen
  \bibfield  {author} {\bibinfo {author} {\bibfnamefont {J.~J.}\ \bibnamefont
  {He}}, \bibinfo {author} {\bibfnamefont {J.}~\bibnamefont {Wu}}, \bibinfo
  {author} {\bibfnamefont {T.-P.}\ \bibnamefont {Choy}}, \bibinfo {author}
  {\bibfnamefont {X.-J.}\ \bibnamefont {Liu}}, \bibinfo {author} {\bibfnamefont
  {Y.}~\bibnamefont {Tanaka}}, \ and\ \bibinfo {author} {\bibfnamefont {K.~T.}\
  \bibnamefont {Law}},\ }\href@noop {} {\bibfield  {journal} {\bibinfo
  {journal} {Nature Communications}\ }\textbf {\bibinfo {volume} {5}},\
  \bibinfo {pages} {3232} (\bibinfo {year} {2014})}\BibitemShut {NoStop}%
\bibitem [{\citenamefont {Xu}\ \emph {et~al.}(2015)\citenamefont {Xu},
  \citenamefont {Wang}, \citenamefont {Liu}, \citenamefont {Ge}, \citenamefont
  {Yang}, \citenamefont {Liu}, \citenamefont {Xu}, \citenamefont {Guan},
  \citenamefont {Gao}, \citenamefont {Qian}, \citenamefont {Liu}, \citenamefont
  {Wang}, \citenamefont {Zhang}, \citenamefont {Xue},\ and\ \citenamefont
  {Jia}}]{xu2015}%
  \BibitemOpen
  \bibfield  {author} {\bibinfo {author} {\bibfnamefont {J.-P.}\ \bibnamefont
  {Xu}}, \bibinfo {author} {\bibfnamefont {M.-X.}\ \bibnamefont {Wang}},
  \bibinfo {author} {\bibfnamefont {Z.~L.}\ \bibnamefont {Liu}}, \bibinfo
  {author} {\bibfnamefont {J.-F.}\ \bibnamefont {Ge}}, \bibinfo {author}
  {\bibfnamefont {X.}~\bibnamefont {Yang}}, \bibinfo {author} {\bibfnamefont
  {C.}~\bibnamefont {Liu}}, \bibinfo {author} {\bibfnamefont {Z.~A.}\
  \bibnamefont {Xu}}, \bibinfo {author} {\bibfnamefont {D.}~\bibnamefont
  {Guan}}, \bibinfo {author} {\bibfnamefont {C.~L.}\ \bibnamefont {Gao}},
  \bibinfo {author} {\bibfnamefont {D.}~\bibnamefont {Qian}}, \bibinfo {author}
  {\bibfnamefont {Y.}~\bibnamefont {Liu}}, \bibinfo {author} {\bibfnamefont
  {Q.-H.}\ \bibnamefont {Wang}}, \bibinfo {author} {\bibfnamefont {F.-C.}\
  \bibnamefont {Zhang}}, \bibinfo {author} {\bibfnamefont {Q.-K.}\ \bibnamefont
  {Xue}}, \ and\ \bibinfo {author} {\bibfnamefont {J.-F.}\ \bibnamefont
  {Jia}},\ }\href {\doibase 10.1103/PhysRevLett.114.017001} {\bibfield
  {journal} {\bibinfo  {journal} {Phys. Rev. Lett.}\ }\textbf {\bibinfo
  {volume} {114}},\ \bibinfo {pages} {017001} (\bibinfo {year}
  {2015})}\BibitemShut {NoStop}%
\bibitem [{\citenamefont {Qi}\ \emph {et~al.}(2010)\citenamefont {Qi},
  \citenamefont {Hughes},\ and\ \citenamefont {Zhang}}]{qi2010b}%
  \BibitemOpen
  \bibfield  {author} {\bibinfo {author} {\bibfnamefont {X.-L.}\ \bibnamefont
  {Qi}}, \bibinfo {author} {\bibfnamefont {T.~L.}\ \bibnamefont {Hughes}}, \
  and\ \bibinfo {author} {\bibfnamefont {S.-C.}\ \bibnamefont {Zhang}},\ }\href
  {\doibase 10.1103/PhysRevB.82.184516} {\bibfield  {journal} {\bibinfo
  {journal} {Phys. Rev. B}\ }\textbf {\bibinfo {volume} {82}},\ \bibinfo
  {pages} {184516} (\bibinfo {year} {2010})}\BibitemShut {NoStop}%
\bibitem [{\citenamefont {Chung}\ \emph {et~al.}(2011)\citenamefont {Chung},
  \citenamefont {Qi}, \citenamefont {Maciejko},\ and\ \citenamefont
  {Zhang}}]{chung2011}%
  \BibitemOpen
  \bibfield  {author} {\bibinfo {author} {\bibfnamefont {S.~B.}\ \bibnamefont
  {Chung}}, \bibinfo {author} {\bibfnamefont {X.-L.}\ \bibnamefont {Qi}},
  \bibinfo {author} {\bibfnamefont {J.}~\bibnamefont {Maciejko}}, \ and\
  \bibinfo {author} {\bibfnamefont {S.-C.}\ \bibnamefont {Zhang}},\ }\href
  {\doibase 10.1103/PhysRevB.83.100512} {\bibfield  {journal} {\bibinfo
  {journal} {Phys. Rev. B}\ }\textbf {\bibinfo {volume} {83}},\ \bibinfo
  {pages} {100512} (\bibinfo {year} {2011})}\BibitemShut {NoStop}%
\bibitem [{\citenamefont {Wang}\ \emph {et~al.}(2015)\citenamefont {Wang},
  \citenamefont {Zhou}, \citenamefont {Lian},\ and\ \citenamefont
  {Zhang}}]{wang2015a}%
  \BibitemOpen
  \bibfield  {author} {\bibinfo {author} {\bibfnamefont {J.}~\bibnamefont
  {Wang}}, \bibinfo {author} {\bibfnamefont {Q.}~\bibnamefont {Zhou}}, \bibinfo
  {author} {\bibfnamefont {B.}~\bibnamefont {Lian}}, \ and\ \bibinfo {author}
  {\bibfnamefont {S.-C.}\ \bibnamefont {Zhang}},\ }\href {\doibase
  10.1103/PhysRevB.92.064520} {\bibfield  {journal} {\bibinfo  {journal} {Phys.
  Rev. B}\ }\textbf {\bibinfo {volume} {92}},\ \bibinfo {pages} {064520}
  (\bibinfo {year} {2015})}\BibitemShut {NoStop}%
\bibitem [{\citenamefont {He}\ \emph {et~al.}(2017)\citenamefont {He},
  \citenamefont {Pan}, \citenamefont {Stern}, \citenamefont {Burks},
  \citenamefont {Che}, \citenamefont {Yin}, \citenamefont {Wang}, \citenamefont
  {Lian}, \citenamefont {Zhou}, \citenamefont {Choi}, \citenamefont {Murata},
  \citenamefont {Kou}, \citenamefont {Chen}, \citenamefont {Nie}, \citenamefont
  {Shao}, \citenamefont {Fan}, \citenamefont {Zhang}, \citenamefont {Liu},
  \citenamefont {Xia},\ and\ \citenamefont {Wang}}]{he2017}%
  \BibitemOpen
  \bibfield  {author} {\bibinfo {author} {\bibfnamefont {Q.~L.}\ \bibnamefont
  {He}}, \bibinfo {author} {\bibfnamefont {L.}~\bibnamefont {Pan}}, \bibinfo
  {author} {\bibfnamefont {A.~L.}\ \bibnamefont {Stern}}, \bibinfo {author}
  {\bibfnamefont {E.~C.}\ \bibnamefont {Burks}}, \bibinfo {author}
  {\bibfnamefont {X.}~\bibnamefont {Che}}, \bibinfo {author} {\bibfnamefont
  {G.}~\bibnamefont {Yin}}, \bibinfo {author} {\bibfnamefont {J.}~\bibnamefont
  {Wang}}, \bibinfo {author} {\bibfnamefont {B.}~\bibnamefont {Lian}}, \bibinfo
  {author} {\bibfnamefont {Q.}~\bibnamefont {Zhou}}, \bibinfo {author}
  {\bibfnamefont {E.~S.}\ \bibnamefont {Choi}}, \bibinfo {author}
  {\bibfnamefont {K.}~\bibnamefont {Murata}}, \bibinfo {author} {\bibfnamefont
  {X.}~\bibnamefont {Kou}}, \bibinfo {author} {\bibfnamefont {Z.}~\bibnamefont
  {Chen}}, \bibinfo {author} {\bibfnamefont {T.}~\bibnamefont {Nie}}, \bibinfo
  {author} {\bibfnamefont {Q.}~\bibnamefont {Shao}}, \bibinfo {author}
  {\bibfnamefont {Y.}~\bibnamefont {Fan}}, \bibinfo {author} {\bibfnamefont
  {S.-C.}\ \bibnamefont {Zhang}}, \bibinfo {author} {\bibfnamefont
  {K.}~\bibnamefont {Liu}}, \bibinfo {author} {\bibfnamefont {J.}~\bibnamefont
  {Xia}}, \ and\ \bibinfo {author} {\bibfnamefont {K.~L.}\ \bibnamefont
  {Wang}},\ }\href {\doibase 10.1126/science.aag2792} {\bibfield  {journal}
  {\bibinfo  {journal} {Science}\ }\textbf {\bibinfo {volume} {357}},\ \bibinfo
  {pages} {294} (\bibinfo {year} {2017})}\BibitemShut {NoStop}%
\bibitem [{\citenamefont {Moore}\ and\ \citenamefont {Read}(1991)}]{moore1991}%
  \BibitemOpen
  \bibfield  {author} {\bibinfo {author} {\bibfnamefont {G.}~\bibnamefont
  {Moore}}\ and\ \bibinfo {author} {\bibfnamefont {N.}~\bibnamefont {Read}},\
  }\href {\doibase http://dx.doi.org/10.1016/0550-3213(91)90407-O} {\bibfield
  {journal} {\bibinfo  {journal} {Nucl. Phys. B}\ }\textbf {\bibinfo {volume}
  {360}},\ \bibinfo {pages} {362 } (\bibinfo {year} {1991})}\BibitemShut
  {NoStop}%
\bibitem [{\citenamefont {Wen}(1991)}]{wen1991}%
  \BibitemOpen
  \bibfield  {author} {\bibinfo {author} {\bibfnamefont {X.~G.}\ \bibnamefont
  {Wen}},\ }\href {\doibase 10.1103/PhysRevLett.66.802} {\bibfield  {journal}
  {\bibinfo  {journal} {Phys. Rev. Lett.}\ }\textbf {\bibinfo {volume} {66}},\
  \bibinfo {pages} {802} (\bibinfo {year} {1991})}\BibitemShut {NoStop}%
\bibitem [{\citenamefont {Read}\ and\ \citenamefont {Green}(2000)}]{read2000}%
  \BibitemOpen
  \bibfield  {author} {\bibinfo {author} {\bibfnamefont {N.}~\bibnamefont
  {Read}}\ and\ \bibinfo {author} {\bibfnamefont {D.}~\bibnamefont {Green}},\
  }\href {\doibase 10.1103/PhysRevB.61.10267} {\bibfield  {journal} {\bibinfo
  {journal} {Phys. Rev. B}\ }\textbf {\bibinfo {volume} {61}},\ \bibinfo
  {pages} {10267} (\bibinfo {year} {2000})}\BibitemShut {NoStop}%
\bibitem [{\citenamefont {Ivanov}(2001)}]{ivanov2001}%
  \BibitemOpen
  \bibfield  {author} {\bibinfo {author} {\bibfnamefont {D.~A.}\ \bibnamefont
  {Ivanov}},\ }\href {\doibase 10.1103/PhysRevLett.86.268} {\bibfield
  {journal} {\bibinfo  {journal} {Phys. Rev. Lett.}\ }\textbf {\bibinfo
  {volume} {86}},\ \bibinfo {pages} {268} (\bibinfo {year} {2001})}\BibitemShut
  {NoStop}%
\bibitem [{\citenamefont {Nayak}\ \emph {et~al.}(2008)\citenamefont {Nayak},
  \citenamefont {Simon}, \citenamefont {Stern}, \citenamefont {Freedman},\ and\
  \citenamefont {Das~Sarma}}]{nayak2008}%
  \BibitemOpen
  \bibfield  {author} {\bibinfo {author} {\bibfnamefont {C.}~\bibnamefont
  {Nayak}}, \bibinfo {author} {\bibfnamefont {S.~H.}\ \bibnamefont {Simon}},
  \bibinfo {author} {\bibfnamefont {A.}~\bibnamefont {Stern}}, \bibinfo
  {author} {\bibfnamefont {M.}~\bibnamefont {Freedman}}, \ and\ \bibinfo
  {author} {\bibfnamefont {S.}~\bibnamefont {Das~Sarma}},\ }\href {\doibase
  10.1103/RevModPhys.80.1083} {\bibfield  {journal} {\bibinfo  {journal} {Rev.
  Mod. Phys.}\ }\textbf {\bibinfo {volume} {80}},\ \bibinfo {pages} {1083}
  (\bibinfo {year} {2008})}\BibitemShut {NoStop}%
\bibitem [{\citenamefont {Kitaev}(2003)}]{kitaev2003}%
  \BibitemOpen
  \bibfield  {author} {\bibinfo {author} {\bibfnamefont {A.}~\bibnamefont
  {Kitaev}},\ }\href {\doibase http://dx.doi.org/10.1016/S0003-4916(02)00018-0}
  {\bibfield  {journal} {\bibinfo  {journal} {Ann. Phys.}\ }\textbf {\bibinfo
  {volume} {303}},\ \bibinfo {pages} {2} (\bibinfo {year} {2003})}\BibitemShut
  {NoStop}%
\bibitem [{\citenamefont {Chang}\ \emph {et~al.}(2013)\citenamefont {Chang},
  \citenamefont {Zhang}, \citenamefont {Feng}, \citenamefont {Shen},
  \citenamefont {Zhang}, \citenamefont {Guo}, \citenamefont {Li}, \citenamefont
  {Ou}, \citenamefont {Wei}, \citenamefont {Wang}, \citenamefont {Ji},
  \citenamefont {Feng}, \citenamefont {Ji}, \citenamefont {Chen}, \citenamefont
  {Jia}, \citenamefont {Dai}, \citenamefont {Fang}, \citenamefont {Zhang},
  \citenamefont {He}, \citenamefont {Wang}, \citenamefont {Lu}, \citenamefont
  {Ma},\ and\ \citenamefont {Xue}}]{chang2013b}%
  \BibitemOpen
  \bibfield  {author} {\bibinfo {author} {\bibfnamefont {C.-Z.}\ \bibnamefont
  {Chang}}, \bibinfo {author} {\bibfnamefont {J.}~\bibnamefont {Zhang}},
  \bibinfo {author} {\bibfnamefont {X.}~\bibnamefont {Feng}}, \bibinfo {author}
  {\bibfnamefont {J.}~\bibnamefont {Shen}}, \bibinfo {author} {\bibfnamefont
  {Z.}~\bibnamefont {Zhang}}, \bibinfo {author} {\bibfnamefont
  {M.}~\bibnamefont {Guo}}, \bibinfo {author} {\bibfnamefont {K.}~\bibnamefont
  {Li}}, \bibinfo {author} {\bibfnamefont {Y.}~\bibnamefont {Ou}}, \bibinfo
  {author} {\bibfnamefont {P.}~\bibnamefont {Wei}}, \bibinfo {author}
  {\bibfnamefont {L.-L.}\ \bibnamefont {Wang}}, \bibinfo {author}
  {\bibfnamefont {Z.-Q.}\ \bibnamefont {Ji}}, \bibinfo {author} {\bibfnamefont
  {Y.}~\bibnamefont {Feng}}, \bibinfo {author} {\bibfnamefont {S.}~\bibnamefont
  {Ji}}, \bibinfo {author} {\bibfnamefont {X.}~\bibnamefont {Chen}}, \bibinfo
  {author} {\bibfnamefont {J.}~\bibnamefont {Jia}}, \bibinfo {author}
  {\bibfnamefont {X.}~\bibnamefont {Dai}}, \bibinfo {author} {\bibfnamefont
  {Z.}~\bibnamefont {Fang}}, \bibinfo {author} {\bibfnamefont {S.-C.}\
  \bibnamefont {Zhang}}, \bibinfo {author} {\bibfnamefont {K.}~\bibnamefont
  {He}}, \bibinfo {author} {\bibfnamefont {Y.}~\bibnamefont {Wang}}, \bibinfo
  {author} {\bibfnamefont {L.}~\bibnamefont {Lu}}, \bibinfo {author}
  {\bibfnamefont {X.-C.}\ \bibnamefont {Ma}}, \ and\ \bibinfo {author}
  {\bibfnamefont {Q.-K.}\ \bibnamefont {Xue}},\ }\href {\doibase
  10.1126/science.1234414} {\bibfield  {journal} {\bibinfo  {journal}
  {Science}\ }\textbf {\bibinfo {volume} {340}},\ \bibinfo {pages} {167}
  (\bibinfo {year} {2013})}\BibitemShut {NoStop}%
\bibitem [{\citenamefont {Checkelsky}\ \emph {et~al.}(2014)\citenamefont
  {Checkelsky}, \citenamefont {Yoshimi}, \citenamefont {Tsukazaki},
  \citenamefont {Takahashi}, \citenamefont {Kozuka}, \citenamefont {Falson},
  \citenamefont {Kawasaki},\ and\ \citenamefont {Tokura}}]{checkelsky2014}%
  \BibitemOpen
  \bibfield  {author} {\bibinfo {author} {\bibfnamefont {J.~G.}\ \bibnamefont
  {Checkelsky}}, \bibinfo {author} {\bibfnamefont {R.}~\bibnamefont {Yoshimi}},
  \bibinfo {author} {\bibfnamefont {A.}~\bibnamefont {Tsukazaki}}, \bibinfo
  {author} {\bibfnamefont {K.~S.}\ \bibnamefont {Takahashi}}, \bibinfo {author}
  {\bibfnamefont {Y.}~\bibnamefont {Kozuka}}, \bibinfo {author} {\bibfnamefont
  {J.}~\bibnamefont {Falson}}, \bibinfo {author} {\bibfnamefont
  {M.}~\bibnamefont {Kawasaki}}, \ and\ \bibinfo {author} {\bibfnamefont
  {Y.}~\bibnamefont {Tokura}},\ }\href@noop {} {\bibfield  {journal} {\bibinfo
  {journal} {Nat. Phys.}\ }\textbf {\bibinfo {volume} {10}},\ \bibinfo {pages}
  {731} (\bibinfo {year} {2014})}\BibitemShut {NoStop}%
\bibitem [{\citenamefont {Kou}\ \emph {et~al.}(2014)\citenamefont {Kou},
  \citenamefont {Guo}, \citenamefont {Fan}, \citenamefont {Pan}, \citenamefont
  {Lang}, \citenamefont {Jiang}, \citenamefont {Shao}, \citenamefont {Nie},
  \citenamefont {Murata}, \citenamefont {Tang}, \citenamefont {Wang},
  \citenamefont {He}, \citenamefont {Lee}, \citenamefont {Lee},\ and\
  \citenamefont {Wang}}]{kou2014}%
  \BibitemOpen
  \bibfield  {author} {\bibinfo {author} {\bibfnamefont {X.}~\bibnamefont
  {Kou}}, \bibinfo {author} {\bibfnamefont {S.-T.}\ \bibnamefont {Guo}},
  \bibinfo {author} {\bibfnamefont {Y.}~\bibnamefont {Fan}}, \bibinfo {author}
  {\bibfnamefont {L.}~\bibnamefont {Pan}}, \bibinfo {author} {\bibfnamefont
  {M.}~\bibnamefont {Lang}}, \bibinfo {author} {\bibfnamefont {Y.}~\bibnamefont
  {Jiang}}, \bibinfo {author} {\bibfnamefont {Q.}~\bibnamefont {Shao}},
  \bibinfo {author} {\bibfnamefont {T.}~\bibnamefont {Nie}}, \bibinfo {author}
  {\bibfnamefont {K.}~\bibnamefont {Murata}}, \bibinfo {author} {\bibfnamefont
  {J.}~\bibnamefont {Tang}}, \bibinfo {author} {\bibfnamefont {Y.}~\bibnamefont
  {Wang}}, \bibinfo {author} {\bibfnamefont {L.}~\bibnamefont {He}}, \bibinfo
  {author} {\bibfnamefont {T.-K.}\ \bibnamefont {Lee}}, \bibinfo {author}
  {\bibfnamefont {W.-L.}\ \bibnamefont {Lee}}, \ and\ \bibinfo {author}
  {\bibfnamefont {K.~L.}\ \bibnamefont {Wang}},\ }\href {\doibase
  10.1103/PhysRevLett.113.137201} {\bibfield  {journal} {\bibinfo  {journal}
  {Phys. Rev. Lett.}\ }\textbf {\bibinfo {volume} {113}},\ \bibinfo {pages}
  {137201} (\bibinfo {year} {2014})}\BibitemShut {NoStop}%
\bibitem [{\citenamefont {Chang}\ \emph {et~al.}(2015)\citenamefont {Chang},
  \citenamefont {Zhao}, \citenamefont {Kim}, \citenamefont {Zhang},
  \citenamefont {Assaf}, \citenamefont {Heiman}, \citenamefont {Zhang},
  \citenamefont {Liu}, \citenamefont {Chan},\ and\ \citenamefont
  {Moodera}}]{chang2015}%
  \BibitemOpen
  \bibfield  {author} {\bibinfo {author} {\bibfnamefont {C.-Z.}\ \bibnamefont
  {Chang}}, \bibinfo {author} {\bibfnamefont {W.}~\bibnamefont {Zhao}},
  \bibinfo {author} {\bibfnamefont {D.~Y.}\ \bibnamefont {Kim}}, \bibinfo
  {author} {\bibfnamefont {H.}~\bibnamefont {Zhang}}, \bibinfo {author}
  {\bibfnamefont {B.~A.}\ \bibnamefont {Assaf}}, \bibinfo {author}
  {\bibfnamefont {D.}~\bibnamefont {Heiman}}, \bibinfo {author} {\bibfnamefont
  {S.-C.}\ \bibnamefont {Zhang}}, \bibinfo {author} {\bibfnamefont
  {C.}~\bibnamefont {Liu}}, \bibinfo {author} {\bibfnamefont {M.~H.~W.}\
  \bibnamefont {Chan}}, \ and\ \bibinfo {author} {\bibfnamefont {J.~S.}\
  \bibnamefont {Moodera}},\ }\href@noop {} {\bibfield  {journal} {\bibinfo
  {journal} {Nature Mater.}\ }\textbf {\bibinfo {volume} {14}},\ \bibinfo
  {pages} {473} (\bibinfo {year} {2015})}\BibitemShut {NoStop}%
\bibitem [{ji2()}]{ji2017}%
  \BibitemOpen
  \href@noop {} {}\bibinfo {note} {W. Ji and X.-G. Wen, arXiv: 1708.06214
  (2017).}\BibitemShut {Stop}%
\bibitem [{hua()}]{huang2017}%
  \BibitemOpen
  \href@noop {} {}\bibinfo {note} {Y. Huang, F. Setiawan and J. D. Sau, arXiv:
  1708.06752 (2017).}\BibitemShut {Stop}%
\bibitem [{\citenamefont {Kramer}\ \emph {et~al.}(2005)\citenamefont {Kramer},
  \citenamefont {Ohtsuki},\ and\ \citenamefont {Kettemann}}]{kramer2005}%
  \BibitemOpen
  \bibfield  {author} {\bibinfo {author} {\bibfnamefont {B.}~\bibnamefont
  {Kramer}}, \bibinfo {author} {\bibfnamefont {T.}~\bibnamefont {Ohtsuki}}, \
  and\ \bibinfo {author} {\bibfnamefont {S.}~\bibnamefont {Kettemann}},\ }\href
  {\doibase 10.1016/j.physrep.2005.07.001} {\bibfield  {journal} {\bibinfo
  {journal} {Phys. Rep.}\ }\textbf {\bibinfo {volume} {417}},\ \bibinfo {pages}
  {211} (\bibinfo {year} {2005})}\BibitemShut {NoStop}%
\bibitem [{\citenamefont {Chalker}\ \emph {et~al.}(2001)\citenamefont
  {Chalker}, \citenamefont {Read}, \citenamefont {Kagalovsky}, \citenamefont
  {Horovitz}, \citenamefont {Avishai},\ and\ \citenamefont
  {Ludwig}}]{chalker2001}%
  \BibitemOpen
  \bibfield  {author} {\bibinfo {author} {\bibfnamefont {J.~T.}\ \bibnamefont
  {Chalker}}, \bibinfo {author} {\bibfnamefont {N.}~\bibnamefont {Read}},
  \bibinfo {author} {\bibfnamefont {V.}~\bibnamefont {Kagalovsky}}, \bibinfo
  {author} {\bibfnamefont {B.}~\bibnamefont {Horovitz}}, \bibinfo {author}
  {\bibfnamefont {Y.}~\bibnamefont {Avishai}}, \ and\ \bibinfo {author}
  {\bibfnamefont {A.~W.~W.}\ \bibnamefont {Ludwig}},\ }\href {\doibase
  10.1103/PhysRevB.65.012506} {\bibfield  {journal} {\bibinfo  {journal} {Phys.
  Rev. B}\ }\textbf {\bibinfo {volume} {65}},\ \bibinfo {pages} {012506}
  (\bibinfo {year} {2001})}\BibitemShut {NoStop}%
\bibitem [{\citenamefont {Anantram}\ and\ \citenamefont
  {Datta}(1996)}]{anantram1996}%
  \BibitemOpen
  \bibfield  {author} {\bibinfo {author} {\bibfnamefont {M.~P.}\ \bibnamefont
  {Anantram}}\ and\ \bibinfo {author} {\bibfnamefont {S.}~\bibnamefont
  {Datta}},\ }\href {\doibase 10.1103/PhysRevB.53.16390} {\bibfield  {journal}
  {\bibinfo  {journal} {Phys. Rev. B}\ }\textbf {\bibinfo {volume} {53}},\
  \bibinfo {pages} {16390} (\bibinfo {year} {1996})}\BibitemShut {NoStop}%
\bibitem [{\citenamefont {Entin-Wohlman}\ \emph {et~al.}(2008)\citenamefont
  {Entin-Wohlman}, \citenamefont {Imry},\ and\ \citenamefont
  {Aharony}}]{entin2008}%
  \BibitemOpen
  \bibfield  {author} {\bibinfo {author} {\bibfnamefont {O.}~\bibnamefont
  {Entin-Wohlman}}, \bibinfo {author} {\bibfnamefont {Y.}~\bibnamefont {Imry}},
  \ and\ \bibinfo {author} {\bibfnamefont {A.}~\bibnamefont {Aharony}},\ }\href
  {\doibase 10.1103/PhysRevB.78.224510} {\bibfield  {journal} {\bibinfo
  {journal} {Phys. Rev. B}\ }\textbf {\bibinfo {volume} {78}},\ \bibinfo
  {pages} {224510} (\bibinfo {year} {2008})}\BibitemShut {NoStop}%
\bibitem [{\citenamefont {Chalker}\ and\ \citenamefont
  {Coddington}(1988)}]{chalker1988}%
  \BibitemOpen
  \bibfield  {author} {\bibinfo {author} {\bibfnamefont {J.~T.}\ \bibnamefont
  {Chalker}}\ and\ \bibinfo {author} {\bibfnamefont {P.~D.}\ \bibnamefont
  {Coddington}},\ }\href@noop {} {\bibfield  {journal} {\bibinfo  {journal} {J.
  Phys. C}\ }\textbf {\bibinfo {volume} {21}},\ \bibinfo {pages} {2665}
  (\bibinfo {year} {1988})}\BibitemShut {NoStop}%
\bibitem [{\citenamefont {Wang}\ \emph {et~al.}(2014)\citenamefont {Wang},
  \citenamefont {Lian},\ and\ \citenamefont {Zhang}}]{wang2014a}%
  \BibitemOpen
  \bibfield  {author} {\bibinfo {author} {\bibfnamefont {J.}~\bibnamefont
  {Wang}}, \bibinfo {author} {\bibfnamefont {B.}~\bibnamefont {Lian}}, \ and\
  \bibinfo {author} {\bibfnamefont {S.-C.}\ \bibnamefont {Zhang}},\ }\href
  {\doibase 10.1103/PhysRevB.89.085106} {\bibfield  {journal} {\bibinfo
  {journal} {Phys. Rev. B}\ }\textbf {\bibinfo {volume} {89}},\ \bibinfo
  {pages} {085106} (\bibinfo {year} {2014})}\BibitemShut {NoStop}%
\bibitem [{\citenamefont {Fulga}\ \emph {et~al.}(2012)\citenamefont {Fulga},
  \citenamefont {Akhmerov}, \citenamefont {Tworzyd\l{}o}, \citenamefont
  {B\'eri},\ and\ \citenamefont {Beenakker}}]{fulga2012}%
  \BibitemOpen
  \bibfield  {author} {\bibinfo {author} {\bibfnamefont {I.~C.}\ \bibnamefont
  {Fulga}}, \bibinfo {author} {\bibfnamefont {A.~R.}\ \bibnamefont {Akhmerov}},
  \bibinfo {author} {\bibfnamefont {J.}~\bibnamefont {Tworzyd\l{}o}}, \bibinfo
  {author} {\bibfnamefont {B.}~\bibnamefont {B\'eri}}, \ and\ \bibinfo {author}
  {\bibfnamefont {C.~W.~J.}\ \bibnamefont {Beenakker}},\ }\href {\doibase
  10.1103/PhysRevB.86.054505} {\bibfield  {journal} {\bibinfo  {journal} {Phys.
  Rev. B}\ }\textbf {\bibinfo {volume} {86}},\ \bibinfo {pages} {054505}
  (\bibinfo {year} {2012})}\BibitemShut {NoStop}%
\bibitem [{\citenamefont {Pruisken}(1984)}]{pruisken1984}%
  \BibitemOpen
  \bibfield  {author} {\bibinfo {author} {\bibfnamefont {A.}~\bibnamefont
  {Pruisken}},\ }\href {\doibase 10.1016/0550-3213(84)90101-9} {\bibfield
  {journal} {\bibinfo  {journal} {Nucl. Phys. B}\ }\textbf {\bibinfo {volume}
  {235}},\ \bibinfo {pages} {277} (\bibinfo {year} {1984})}\BibitemShut
  {NoStop}%
\bibitem [{\citenamefont {Pruisken}(1988)}]{pruisken1988}%
  \BibitemOpen
  \bibfield  {author} {\bibinfo {author} {\bibfnamefont {A.~M.~M.}\
  \bibnamefont {Pruisken}},\ }\href {\doibase 10.1103/PhysRevLett.61.1297}
  {\bibfield  {journal} {\bibinfo  {journal} {Phys. Rev. Lett.}\ }\textbf
  {\bibinfo {volume} {61}},\ \bibinfo {pages} {1297} (\bibinfo {year}
  {1988})}\BibitemShut {NoStop}%
\bibitem [{\citenamefont {Thouless}(1977)}]{thouless1977}%
  \BibitemOpen
  \bibfield  {author} {\bibinfo {author} {\bibfnamefont {D.~J.}\ \bibnamefont
  {Thouless}},\ }\href {\doibase 10.1103/PhysRevLett.39.1167} {\bibfield
  {journal} {\bibinfo  {journal} {Phys. Rev. Lett.}\ }\textbf {\bibinfo
  {volume} {39}},\ \bibinfo {pages} {1167} (\bibinfo {year}
  {1977})}\BibitemShut {NoStop}%
\bibitem [{\citenamefont {Sondhi}\ \emph {et~al.}(1997)\citenamefont {Sondhi},
  \citenamefont {Girvin}, \citenamefont {Carini},\ and\ \citenamefont
  {Shahar}}]{sondhi1997}%
  \BibitemOpen
  \bibfield  {author} {\bibinfo {author} {\bibfnamefont {S.~L.}\ \bibnamefont
  {Sondhi}}, \bibinfo {author} {\bibfnamefont {S.~M.}\ \bibnamefont {Girvin}},
  \bibinfo {author} {\bibfnamefont {J.~P.}\ \bibnamefont {Carini}}, \ and\
  \bibinfo {author} {\bibfnamefont {D.}~\bibnamefont {Shahar}},\ }\href
  {\doibase 10.1103/RevModPhys.69.315} {\bibfield  {journal} {\bibinfo
  {journal} {Rev. Mod. Phys.}\ }\textbf {\bibinfo {volume} {69}},\ \bibinfo
  {pages} {315} (\bibinfo {year} {1997})}\BibitemShut {NoStop}%
\bibitem [{\citenamefont {Huckestein}\ and\ \citenamefont
  {Kramer}(1990)}]{huckestein1990}%
  \BibitemOpen
  \bibfield  {author} {\bibinfo {author} {\bibfnamefont {B.}~\bibnamefont
  {Huckestein}}\ and\ \bibinfo {author} {\bibfnamefont {B.}~\bibnamefont
  {Kramer}},\ }\href {\doibase 10.1103/PhysRevLett.64.1437} {\bibfield
  {journal} {\bibinfo  {journal} {Phys. Rev. Lett.}\ }\textbf {\bibinfo
  {volume} {64}},\ \bibinfo {pages} {1437} (\bibinfo {year}
  {1990})}\BibitemShut {NoStop}%
\bibitem [{\citenamefont {Kou}\ \emph {et~al.}(2015)\citenamefont {Kou},
  \citenamefont {Pan}, \citenamefont {Wang}, \citenamefont {Fan}, \citenamefont
  {Choi}, \citenamefont {Lee}, \citenamefont {Nie}, \citenamefont {Murata},
  \citenamefont {Shao}, \citenamefont {Zhang},\ and\ \citenamefont
  {Wang}}]{kou2015}%
  \BibitemOpen
  \bibfield  {author} {\bibinfo {author} {\bibfnamefont {X.}~\bibnamefont
  {Kou}}, \bibinfo {author} {\bibfnamefont {L.}~\bibnamefont {Pan}}, \bibinfo
  {author} {\bibfnamefont {J.}~\bibnamefont {Wang}}, \bibinfo {author}
  {\bibfnamefont {Y.}~\bibnamefont {Fan}}, \bibinfo {author} {\bibfnamefont
  {E.~S.}\ \bibnamefont {Choi}}, \bibinfo {author} {\bibfnamefont {W.-L.}\
  \bibnamefont {Lee}}, \bibinfo {author} {\bibfnamefont {T.}~\bibnamefont
  {Nie}}, \bibinfo {author} {\bibfnamefont {K.}~\bibnamefont {Murata}},
  \bibinfo {author} {\bibfnamefont {Q.}~\bibnamefont {Shao}}, \bibinfo {author}
  {\bibfnamefont {S.-C.}\ \bibnamefont {Zhang}}, \ and\ \bibinfo {author}
  {\bibfnamefont {K.~L.}\ \bibnamefont {Wang}},\ }\href {\doibase
  10.1038/ncomms9474} {\bibfield  {journal} {\bibinfo  {journal} {Nat.
  Commun.}\ }\textbf {\bibinfo {volume} {6}},\ \bibinfo {pages} {8474}
  (\bibinfo {year} {2015})}\BibitemShut {NoStop}%
\bibitem [{\citenamefont {Biscaras}\ \emph {et~al.}(2013)\citenamefont
  {Biscaras}, \citenamefont {Bergeal}, \citenamefont {Hurand}, \citenamefont
  {Feuillet-Palma}, \citenamefont {Rastogi}, \citenamefont {Budhani},
  \citenamefont {Grilli}, \citenamefont {Caprara},\ and\ \citenamefont
  {Lesueur}}]{biscaras2013}%
  \BibitemOpen
  \bibfield  {author} {\bibinfo {author} {\bibfnamefont {J.}~\bibnamefont
  {Biscaras}}, \bibinfo {author} {\bibfnamefont {N.}~\bibnamefont {Bergeal}},
  \bibinfo {author} {\bibfnamefont {S.}~\bibnamefont {Hurand}}, \bibinfo
  {author} {\bibfnamefont {C.}~\bibnamefont {Feuillet-Palma}}, \bibinfo
  {author} {\bibfnamefont {A.}~\bibnamefont {Rastogi}}, \bibinfo {author}
  {\bibfnamefont {R.~C.}\ \bibnamefont {Budhani}}, \bibinfo {author}
  {\bibfnamefont {M.}~\bibnamefont {Grilli}}, \bibinfo {author} {\bibfnamefont
  {S.}~\bibnamefont {Caprara}}, \ and\ \bibinfo {author} {\bibfnamefont
  {J.}~\bibnamefont {Lesueur}},\ }\href@noop {} {\bibfield  {journal} {\bibinfo
   {journal} {Nat. Mater.}\ }\textbf {\bibinfo {volume} {12}},\ \bibinfo
  {pages} {542} (\bibinfo {year} {2013})}\BibitemShut {NoStop}%
\bibitem [{\citenamefont {Lachman}\ \emph {et~al.}(2015)\citenamefont
  {Lachman}, \citenamefont {Young}, \citenamefont {Richardella}, \citenamefont
  {Cuppens}, \citenamefont {Naren}, \citenamefont {Anahory}, \citenamefont
  {Meltzer}, \citenamefont {Kandala}, \citenamefont {Kempinger}, \citenamefont
  {Myasoedov}, \citenamefont {Huber}, \citenamefont {Samarth},\ and\
  \citenamefont {Zeldov}}]{lachmane2015}%
  \BibitemOpen
  \bibfield  {author} {\bibinfo {author} {\bibfnamefont {E.~O.}\ \bibnamefont
  {Lachman}}, \bibinfo {author} {\bibfnamefont {A.~F.}\ \bibnamefont {Young}},
  \bibinfo {author} {\bibfnamefont {A.}~\bibnamefont {Richardella}}, \bibinfo
  {author} {\bibfnamefont {J.}~\bibnamefont {Cuppens}}, \bibinfo {author}
  {\bibfnamefont {H.~R.}\ \bibnamefont {Naren}}, \bibinfo {author}
  {\bibfnamefont {Y.}~\bibnamefont {Anahory}}, \bibinfo {author} {\bibfnamefont
  {A.~Y.}\ \bibnamefont {Meltzer}}, \bibinfo {author} {\bibfnamefont
  {A.}~\bibnamefont {Kandala}}, \bibinfo {author} {\bibfnamefont
  {S.}~\bibnamefont {Kempinger}}, \bibinfo {author} {\bibfnamefont
  {Y.}~\bibnamefont {Myasoedov}}, \bibinfo {author} {\bibfnamefont {M.~E.}\
  \bibnamefont {Huber}}, \bibinfo {author} {\bibfnamefont {N.}~\bibnamefont
  {Samarth}}, \ and\ \bibinfo {author} {\bibfnamefont {E.}~\bibnamefont
  {Zeldov}},\ }\href {\doibase 10.1126/sciadv.1500740} {\bibfield  {journal}
  {\bibinfo  {journal} {Science Advances}\ }\textbf {\bibinfo {volume} {1}}
  (\bibinfo {year} {2015}),\ 10.1126/sciadv.1500740}\BibitemShut {NoStop}%
\bibitem [{\citenamefont {Senthil}\ and\ \citenamefont
  {Fisher}(2000)}]{senthil2000}%
  \BibitemOpen
  \bibfield  {author} {\bibinfo {author} {\bibfnamefont {T.}~\bibnamefont
  {Senthil}}\ and\ \bibinfo {author} {\bibfnamefont {M.~P.~A.}\ \bibnamefont
  {Fisher}},\ }\href {\doibase 10.1103/PhysRevB.61.9690} {\bibfield  {journal}
  {\bibinfo  {journal} {Phys. Rev. B}\ }\textbf {\bibinfo {volume} {61}},\
  \bibinfo {pages} {9690} (\bibinfo {year} {2000})}\BibitemShut {NoStop}%
\bibitem [{\citenamefont {Medvedyeva}\ \emph {et~al.}(2010)\citenamefont
  {Medvedyeva}, \citenamefont {Tworzyd\l{}o},\ and\ \citenamefont
  {Beenakker}}]{medvedyeva2010}%
  \BibitemOpen
  \bibfield  {author} {\bibinfo {author} {\bibfnamefont {M.~V.}\ \bibnamefont
  {Medvedyeva}}, \bibinfo {author} {\bibfnamefont {J.}~\bibnamefont
  {Tworzyd\l{}o}}, \ and\ \bibinfo {author} {\bibfnamefont {C.~W.~J.}\
  \bibnamefont {Beenakker}},\ }\href {\doibase 10.1103/PhysRevB.81.214203}
  {\bibfield  {journal} {\bibinfo  {journal} {Phys. Rev. B}\ }\textbf {\bibinfo
  {volume} {81}},\ \bibinfo {pages} {214203} (\bibinfo {year}
  {2010})}\BibitemShut {NoStop}%
\bibitem [{\citenamefont {Wimmer}\ \emph {et~al.}(2010)\citenamefont {Wimmer},
  \citenamefont {Akhmerov}, \citenamefont {Medvedyeva}, \citenamefont
  {Tworzyd\l{}o},\ and\ \citenamefont {Beenakker}}]{wimmer2010}%
  \BibitemOpen
  \bibfield  {author} {\bibinfo {author} {\bibfnamefont {M.}~\bibnamefont
  {Wimmer}}, \bibinfo {author} {\bibfnamefont {A.~R.}\ \bibnamefont
  {Akhmerov}}, \bibinfo {author} {\bibfnamefont {M.~V.}\ \bibnamefont
  {Medvedyeva}}, \bibinfo {author} {\bibfnamefont {J.}~\bibnamefont
  {Tworzyd\l{}o}}, \ and\ \bibinfo {author} {\bibfnamefont {C.~W.~J.}\
  \bibnamefont {Beenakker}},\ }\href {\doibase 10.1103/PhysRevLett.105.046803}
  {\bibfield  {journal} {\bibinfo  {journal} {Phys. Rev. Lett.}\ }\textbf
  {\bibinfo {volume} {105}},\ \bibinfo {pages} {046803} (\bibinfo {year}
  {2010})}\BibitemShut {NoStop}%
\bibitem [{\citenamefont {Liu}\ \emph {et~al.}(2010)\citenamefont {Liu},
  \citenamefont {Qi}, \citenamefont {Zhang}, \citenamefont {Dai}, \citenamefont
  {Fang},\ and\ \citenamefont {Zhang}}]{liu2010}%
  \BibitemOpen
  \bibfield  {author} {\bibinfo {author} {\bibfnamefont {C.-X.}\ \bibnamefont
  {Liu}}, \bibinfo {author} {\bibfnamefont {X.-L.}\ \bibnamefont {Qi}},
  \bibinfo {author} {\bibfnamefont {H.}~\bibnamefont {Zhang}}, \bibinfo
  {author} {\bibfnamefont {X.}~\bibnamefont {Dai}}, \bibinfo {author}
  {\bibfnamefont {Z.}~\bibnamefont {Fang}}, \ and\ \bibinfo {author}
  {\bibfnamefont {S.-C.}\ \bibnamefont {Zhang}},\ }\href {\doibase
  10.1103/PhysRevB.82.045122} {\bibfield  {journal} {\bibinfo  {journal} {Phys.
  Rev. B}\ }\textbf {\bibinfo {volume} {82}},\ \bibinfo {pages} {045122}
  (\bibinfo {year} {2010})}\BibitemShut {NoStop}%
\end{thebibliography}

%

\end{document}